\documentclass[aps,nofootinbib,prd,class-pre,showpacs]{revtex4} 
\usepackage{amsmath,amssymb,amsfonts,amsthm}
\usepackage{graphicx}

\usepackage{subeqnarray}

\newcommand{\be}{\begin{equation}}
\newcommand{\ee}{\end{equation}}
\newcommand{\beq}{\begin{eqnarray}}
\newcommand{\eeq}{\end{eqnarray}}

\newcommand{\bsa}{\begin{subeqnarray}}
\newcommand{\esa}{\end{subeqnarray}}

\def\lsim{\hbox{ \raise.35ex\rlap{$<$}\lower.6ex\hbox{$\sim$}\ }}
\def\gsim{\hbox{ \raise.35ex\rlap{$>$}\lower.6ex\hbox{$\sim$}\ }}

\newcommand{\bea}{\begin{eqnarray}}
\newcommand{\eea}{\end{eqnarray}}
\newcommand{\beaa}{\begin{eqnarray}}
\newcommand{\eeaa}{\end{eqnarray}}
\newcommand{\ba}{\begin{array}}
\newcommand{\ea}{\end{array}}
\newcommand{\bit}{\begin{itemize}}
\newcommand{\eit}{\end{itemize}}
\newcommand{\ben}{\begin{enumerate}}
\newcommand{\een}{\end{enumerate}}

\def\lab{\label}

\def\rar{\rightarrow}

\def\al{\alpha}

\def\ga{\gamma}
\def\Ga{\Gamma}

\def\la{\lambda}

\def\si{\sigma}

\def\om{\omega}

\begin{document}
\title{
Cortical phase transitions, non-equilibrium thermodynamics\\ and the time-dependent Ginzburg-Landau equation
}
\author{Walter J. Freeman\footnote{dfreeman@berkeley.edu ~~-~~ http://sulcus.berkeley.edu}}
\affiliation{Department of Molecular and Cell Biology\\
University of California, Berkeley CA 94720-3206 USA}

\author{Roberto Livi\footnote{livi@fi.infn.it}}
\affiliation{Dipartimento di Fisica and Istituto Nazionale di Fisica Nucleare\\ Universit\'a di Firenze,  I-50022 Sesto Fiorentino (Firenze), Italy}

\author{Masashi Obinata\footnote{mobinata@unisa.it}}
\affiliation{Facolt\'a di Scienze and Istituto Nazionale di Fisica Nucleare\\ Universit\'a di Salerno, I-84100 Fisciano (Salerno), Italy\\
and Department of Physics, Tsukuba University, Tsukuba, Japan}

\author{Giuseppe Vitiello\footnote{vitiello@sa.infn.it ~~-~~ www.sa.infn.it/giuseppe.vitiello/}}
\affiliation{Facolt\'a di Scienze and Istituto Nazionale di Fisica Nucleare\\ Universit\'a di Salerno, I-84100 Fisciano (Salerno), Italy}

\begin{abstract}
The formation of amplitude modulated and phase modulated assemblies of neurons is observed in the brain functional activity. The study of the formation of such structures requires that the analysis has to be organized in  hierarchical levels, microscopic, mesoscopic, macroscopic, each with its characteristic space-time scales and the various forms of energy, electric, chemical, thermal produced and used by the brain. In this paper, we discuss the microscopic dynamics underlying the mesoscopic and the macroscopic levels and focus our attention on the thermodynamics of the non-equilibrium phase transitions. We obtain the time-dependent Ginzburg-Landau equation for the non-stationary regime and consider the formation of topologically non-trivial structures such as the vortex solution. The power laws observed in functional activities of the brain is also discussed and related to coherent states characterizing the many-body dissipative model of brain.
\end{abstract}

\maketitle

\section{Introduction}

A major effort is under way worldwide to develop superior forms of machine intelligence by applying basic concepts and equations from physics to the interpretation of neurobiological and neuropsychological data. The property that most clearly distinguishes biological intelligence from contemporary machine intelligence is the rich contextualization of information by brains in the construction of knowledge and meaning. Computers and robots operate on information, but they don't know what it means. Books and reprints contain information and display it seriatim in accord with Shannon's theory of information, but knowledge of what the information means is solely in the brains of the writers and readers. We propose that what differentiates knowledge from information are the innumerable linkages among myriad fragments of information, which create the moment-to-moment framework for effective intentional action into the world \cite{Barham}. We find that meaning can best be described objectively as created and carried by great fields of neural activity, which subjectively we experience as thoughts and perceptions. The contents of such fields are constructed from the fragments of information that are imported by sensory neurons and stored by changes in the synaptic linkages among the cortical neurons. It is the dynamics of the populations in each sensory cortex that subsequently organizes the microscopic fragments into meaningful knowledge by creating macroscopic vector fields of activity that organize hundreds of millions of neurons and trillions of synapses. Our aim in this report is to summarize our observations of cortical fields in humans and other animals engaged in intelligent behaviors, and then to describe the dynamics of cortex using concepts adapted from non-equilibrium thermodynamics \cite{Freeman2008} and quantum field theory \cite{11a,Freeman2010}. In particular we focus on adapting the time-dependent Ginzburg-Landau equation so as to describe the construction and transmission of the wave functions in vector fields that we observe in brain activities and experience as knowledge.

A thermodynamic model is needed, because intelligence is extremely energy-intensive. Brains consume free energy in many forms (electric, magnetic, chemical, metabolic) at rates ten-fold greater than any other organ. Salient among these forms is electric current flowing in closed loops that is carried by ions (not electrons) in water. The flux of ions within neurons
is essential for their functions: to communicate by transmitting pulses with axons, and to sum synaptic potentials with dendrites \cite{Freeman2001}. The extracellular flows of loop currents are revealed by ohmic potentials that we record from cortical surfaces (the electrocorticogram, ECoG) and from the scalp (the electroencephalogram, EEG). The cortex is a thin sheet of neurons covering the outer surface of the brain. The average thickness in humans is $\sim 3 ~mm$, and the surface area is $\sim  2,000 ~cm^2$, a ratio of $1:10^5$. Whereas the dynamics of microscopic neurons in networks and the scalar potential fields of the EEG/ECoG are described in 3-D, the macroscopic fields of activity from $10^8$ to $10^9$ neurons each supporting $10^4$ synapses \cite{Braitenberg} topologically operate in 2-D, owing to the long correlation distances compared with the dimensions of neurons.

\begin{figure}
\centering \resizebox{10cm}{!}{\includegraphics{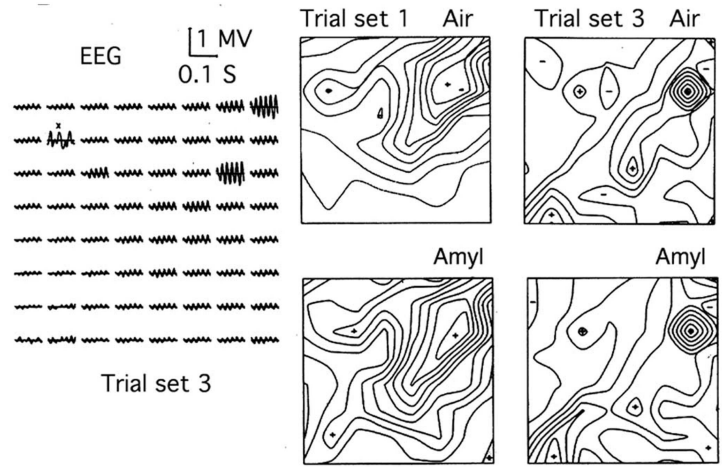}}
\caption{\small \noindent Left: The burst of gamma oscillation illustrates the amplitude modulation of the shared carrier wave. Right: AM patterns are compared with and without the conditioned stimuli (CS) present in the inhaled air. The change between trial sets illustrates consolidation: 'off-line' learning requiring participation of the genome.}
\label{fig1}
\end{figure}

Our standpoint is that cortical dynamics cannot be reduced to the level of neural networks. A macroscopic level of mass action emerges from microscopic dynamics, into which the senses inject their microscopic information and from which macroscopic neural commands are generated. The ECoG and EEG are appropriate to the macroscopic level, because they are summed contributions from interactive masses of neurons. The interactions create dynamic neural assemblies with properties that are undetected with pulses from microelectrodes and inexplicable in terms of neural nets \cite{11}. In the simplest description we conceive cortex as a self-regulating, self-stabilized population of neurons. It modulates and is modulated by other parts of the brain, but it does so on its own terms. Whereas the cortical neural network dynamics involves pulse and wave frequencies, macroscopic population dynamics uses pulse and wave densities, which constitute the state variables in space-time.

In our analysis it is important that we distinguish the level of function by properly defining the state variables. To this aim, since in this paper we are mainly interested in the microscopic dynamics out of which macroscopic structures emerge, we consider the many-body dynamics of the vibrational quanta of the electric dipoles of water molecules. Water is the matrix in which neurons, glial cells and the whole net of dendrites and axons are embedded, including in particular the ionic currents that exercise the electric forces on and in nerve membranes that constitute neural activity. The dipole dynamics provides the possibility for long-range correlation, which in turn allows global synaptic communication among the neurons, each with every other as required for the construction of meaning. It should be clearly stated that the water molecules and their dipole oscillations are not the agency of communication among the neurons. As said above, the memory required for macroscopic structure formation is based in the trillions of synapses between neurons. Likewise, in our prior publications we have made it clear that the extracellular electric fields of the EEG and ECoG are epiphenomenal. They are scalar fields of passive dissipation of electric energy as heat, whereas the fields of neural activity are vector fields, which actively derive free energy from chemical energy in every square micron and dissipate that energy in clouds of action potentials and ultimately as heat. Our many-body analysis shows how such a vector field is indeed generated from the basic dynamics of the dipole vibrational quanta \cite{11,Freeman19}.

The main sources of our experimental data are high-resolution images of the EEG and ECoG from $8 \times 8$ planar arrays of $64$ electrodes \cite{9,10,12,13,Bollobas}. The $64$  signals reveal brief epochs ($\sim 0.1 ~s$) of $3-5$ cycles of narrow band oscillation. The epochs recur $3 -10$ times each second, each time with a different carrier frequency randomly centered in the beta-gamma range. The amplitude of the carrier wave is modulated in a spatial amplitude modulated (AM) pattern (Fig. 1). The set of $64$ amplitudes constitutes a $64 \times 1$ vector that specifies a point in $64$-space. Similar patterns accruing from repeated presentation of a conditioned stimulus (CS) form a cluster of points. Differing CSs form multiple clusters, one for each CS that a subject can discriminate. The AM patterns lack invariance with respect to the CS. The same CS in a different context gives a distinctly different cluster. This contextual dependence shows that the AM patterns are dependent on the meaning of the CSs; they are not representations of the CSs but of the knowledge the subject has about the CSs. The AM patterns evolve slowly with repetition over time and the accumulation of new experiences. Each retrieval of a memory changes the memory by adding new context.

\begin{figure}
\centering \resizebox{10cm}{!}{\includegraphics{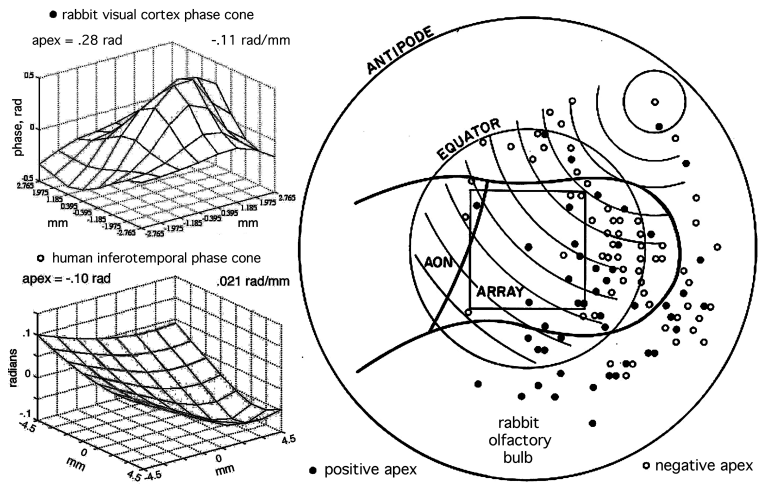}}
\caption{\small \noindent A. An example is shown from rabbit ECoG of a cone fitted to the surface given by an $8 \times 8$ array of values of analytic phase with phase lead at the apex ("explosion"). B. Analytic phase from human ECoG with phase lag at the apex ("implosion").       C. Summary diagram of the variants of phase patterns seen in cinematic displays of the filtered ECoG. Examples can be seen at the URL: http://soma.berkeley.edu}
\label{fig1b}
\end{figure}

Each AM pattern is accompanied by a spatial pattern of phase modulation (PM) of the carrier frequency, which has the form of a cone (Fig. \ref{fig1b}) (see also Appendix \ref{A0}). The location and sign of the apex vary randomly from each epoch to the next. The gradient of the cone in $rad/m$ is fixed in each epoch and varies between bursts inversely with the carrier frequency in $rad/s$. The ratio gives the phase velocity in $m/s$, which is determined by the conduction velocity of pulses on axons running parallel to the cortical surface. We infer that each new AM pattern forms by a state transition, which begins at a site of nucleation on the surface and spreads radially as in the formation of a raindrop. The diameter of the AM pattern is determined by the distance at which phase dispersion by the phase gradient reduces interaction strength below the half-power level.
During an epoch the dynamics is stationary in carrier frequency, AM and PM, and the dynamics is near linear. We introduce the Hilbert transform, which gives the high temporal resolution needed to display the transitions between successive images \cite{9,10}.
We also introduce the concept of criticality, which we need to help explain the capacity of cortex to undergo the dramatic transitions that are required to construct and destroy the images enabling perception \cite{Freeman2008}. Cinematic display of the spatial AM/PM patterns reveals repetitive pulsatile expansions or contractions of the wave functions, and in some instances rotation either clockwise or counterclockwise indicating the stabilization of the AM/PM patterns by formation of vortices, hence vector fields \cite{Freeman2010}.

The point is that the many-body approach allows the formation of spatially delimited domains that constitute the basic facilitating environment wherein the formation occurs of the above described AM and PM structures in the cortex. These domains are also temporally limited in time, in the sense that they have a finite lifetime; they condense and evaporate continually as in a fog, with power-law distributions of duration and diameter. We treat them as the result of a space-time non-homogeneous boson condensation process \cite{Vitiello:1995wv,Vitiello:2001,Alfinito:2002a,Alfinito:2002b} and as neural avalanches by which criticality is maintained \cite{Kitzbichler}.

Intermittently we see the long-lasting epochs generated by the long-range correlation of
neurons locked in coherent oscillation mode carrying AM patterns relating to perception. The dipole quantum field correlation thus acts to facilitate expansion of the trillions of microscopic synapses modified by learning into the spatially textured macroscopic cognition. The expansion is directed by a mesoscopic Hebbian
assembly (Appendix A) into the basin of a macroscopic chaotic attractor that characterizes the AM pattern. The Hebbian assembly provides a burst of transition energy that brains require in order to restrict the expansion to intended, learned sensory input and to exclude environmental noise.  From such a perspective, in the present paper we therefore study the thermodynamics of the phase transition processes at the level of the many-body dynamics of the dipole wave quanta with defined topological properties. The AM/PM patterns then can be considered to be the macroscopic manifestations of a boson condensation and may be characterized by its underlying topological properties. The basic symmetry is the (non-Abelian) rotational SU(2) dipole symmetry, and the emerging structure
formation is entropic, always incurring loss of information particularly in abstraction and inductive generalization. The topology of symmetry breaking from the search/receiving mode to the constructive/transmitting mode is related with the singularity in the null spike where prior structure is expunged and disorder is initiated, as at center (the core) of a vortex \cite{Freeman2010}.

In Section II we present basic elements of the many-body dissipative model with specific attention to the mechanism of boson condensation in a stationary regime. Further details of the formalism of the dissipative many-body model can be found in Refs. \cite{11,Vitiello:1995wv,Vitiello:2001}. The non-stationary regime is discussed in Section III where we obtain the time-dependent Ginzburg-Landau equation. The vortex solution is considered in Section IV and the power laws observed in functional activities of the brain are discussed in Section V and related to coherent states characterizing the many-body dissipative model. Finally, Section VI is devoted to conclusions. Further remarks on the relevant
aspects of the observed brain activity and some mathematical details are presented in the Appendices.

\section{The dissipative many-body model of brain}\label{II}

In the dissipative many-body model of brain \cite{Vitiello:1995wv,Vitiello:2001}, one considers the spontaneous breakdown of the rotational symmetry of the dipoles of water molecules. In quantum field theory (QFT) spontaneous breakdown of symmetry produces the formation of ordered patterns through the mechanism of boson condensation.
Such ordered patterns, resulting into ``in phase" dipole oscillations and often also space ordering  are the manifestation of coherent dynamic features, out of which collective macroscopic behaviors of coherent spatial domains appear \cite{Umezawa:1993yq,DifettiBook}. The lagrangian describing the dynamics of the
system is assumed to be invariant under the SU(2) group of rotations for the
molecular dipole vibrational field $\phi(x, t) = (\phi_\uparrow (x, t) ~~~~ \phi_\downarrow (x, t))^T$ ($j = \uparrow ,\downarrow $, denotes the dipole quantum number and
assumes, for simplicity, only two values "up" and "down") \cite{DelGiudice:1985,DelGiudice:1986}. On the contrary, the
ground state (the "vacuum") of the system cannot be SU(2) invariant since it has to
represent the electret state of water. The SU(2) spontaneous breakdown is expressed
by the relation
\be\lab{2.1}
\langle 0| D^{(3)} ({\bf r}, t) |0 \rangle = {\cal P}({\bf r}, t) \neq O,
\ee
where $D^{(3)} ({\bf r}, t)$ denotes the dipole density operator in the third direction and  ${\cal P}({\bf r}, t)$ is the polarization density. The dipole density operators $D^{(i)} ({\bf r}, t)$, $i = 1,2, 3$, are functionals
of the molecular vibrational field $\phi(x, t)$.
For example, they can be expressed as
$D^{(i)} ({\bf r}, t) = \phi^{\dag}(x, t) \frac{1}{2} \si_i \phi(x, t)$, $i=1,2,3$,
with $\si$ the Pauli matrices. The conclusions
we will reach, however,  are not dependent on the particular expression of
$D^{(i)} ({\bf r}, t)$  in terms of $\phi(x, t)$. A general theorem in quantum field theory (QFT) \cite{Goldstone:1961eq,ITZ,Umezawa:1982nv,Umezawa:1993yq,DifettiBook} shows that spontaneous breakdown of symmetry
implies the existence of gapless fields (the Nambu-Goldstone modes or particles (NG)). They appear as
{\it collective} excitations describing the long-range correlation among the dipoles generating the ordering of the system (electret state). In the dissipative many-body model of brain, a number of functional features of brain can be understood as manifestations
of the NG collective modes \cite{Vitiello:1995wv,Vitiello:2001}.

The  spontaneous breakdown of SU(2) symmetry represented by Eq.~(\ref{2.1}) means that the vacuum is not invariant under rotations induced by $D^{(1)} ({\bf r}, t)$ and $D^{(2)} ({\bf r}, t)$. It is, however, invariant under rotations induced by $D^{(3)} ({\bf r}, t)$, i.e. under the U(1) rotation group around the third direction:
\be \label{u1transf}
\phi \rightarrow \exp(iq\la_{3} ({\bf r}, t)\si_{3}/2)\phi ~.
\ee
Here  $q$ is the electric charge at the head of the molecular dipole. We assume $\la_{3} ({\bf r}, t) \rar 0$ for $|t|\rar \infty$ and/or $|{\bf r}|\rar
\infty$.

Of course, invariance of the lagrangian under
local (i.e. with space-time-dependent  $\la_3 ({\bf r}, t)$) U(1) rotation group signals the presence (requires the introduction) of the electromagnetic interaction
among the dipoles, i.e.  the electromagnetic vector potential
$A_\mu ({\bf r}, t)$, which transforms as
\be\lab{2.5b}
A_\mu ({\bf r}, t) \rightarrow A_\mu ({\bf r}, t) - \partial_\mu \la_3 ({\bf r}, t) ,
\ee
when the phase transformation around the third axis is performed. In the following, we will use the Coulomb gauge ${\mbox{\boldmath $\nabla$}}\cdot {\rm\bf A} = 0$ as gauge condition.
In terms of the dipole operators, the local  U(1) transformation is
\be\lab{2.7}
D^{(\pm)} ({\bf r}, t) \rightarrow  \exp(\mp iq\la_{3} ({\bf r}, t) ) D^{(\pm)} ,
\ee
where
\be\lab{2.6}
D^{(\pm)} ({\bf r}, t) = D^{(1)}({\bf r}, t)  \pm i D^{(2)}({\bf r}, t)  .
\ee

Next we observe that actually the ground state of the system  does not exhibit invariance under global (i.e. space-time independent $\la_3$) U(1) symmetry; this would
require the possibility of changing the phase of the molecular vibrational
field $\phi ({\bf r}, t)$ simultaneously at every space-point by the constant amount $\la_3$. Therefore, the global U(1) symmetry is also spontaneously broken. The condition which expresses the symmetry breakdown is the non-vanishing expectation value of $D^{(+)} ({\bf r}, t)$ in the ground state $|0 \rangle$:
\be\lab{2.9}
\langle 0| D^{(+)} ({\bf r}, t) |0 \rangle = v ({\bf r}, t) \neq O,
\ee
and its complex conjugate relation involving $D^{(-)} ({\bf r}, t)$. $v ({\bf r}, t)$ is a complex function and is called the {\it order parameter}, which is thus a (classical) vector field. It describes macroscopic collective properties of the system. Its space-time dependence denotes (space-time) non-homogeneities in the ground state.  Since the vector potential $A_{\mu}$ is also involved in the dynamics, general theorems of QFT  predict some dynamical effects globally named as the Anderson-Higgs-Kibble mechanism \cite{hig,Anderson:1984a}. Here, we will not consider such a mechanism (see Refs. \cite{ITZ,Umezawa:1982nv,Umezawa:1993yq,DifettiBook} for a general discussion).

Let $\widehat{P} ({\bf r}, t)$ and $\widehat{P}^{\dag} ({\bf r}, t)$ denote the
annihilation and the creation operator, respectively, of the dipole-wave quantum
(the NG mode implied by the spontaneous breakdown of SU(2) symmetry).
By resorting to the results of Refs. \cite{DelGiudice:1985,DelGiudice:1986}, we have that the ground state turns out to be  eigenvector of the dipole-wave operator $\widehat{P} ({\bf r}, t)$:
\be\lab{2.12}
\widehat{P} ({\bf r}, t) |0 \rangle =  \frac{v ({\bf r}, t)}{(2{\cal P})^{1/2}} |0 \rangle ,
\ee
which shows that the ground state is a {\it coherent state}: dipole-wave quanta (the NG boson modes) are coherently condensate in the state $|0 \rangle$  \cite{DelGiudice:1985,DelGiudice:1986}. We also have
\be\lab{2.13}
\langle 0| D^{(-)}({\bf r}, t) D^{(+)} ({\bf r}, t) |0 \rangle = 2{\cal P} ({\bf r}, t) \,  \langle 0| \widehat{P}^{\dag} ({\bf r}, t)  \widehat{P} ({\bf r}, t) |0 \rangle = |v ({\bf r}, t)|^2 , \ee
We may also define the polarization density as
\be\lab{2.16}
 \langle 0| D^{(-)}({\bf r}, t) D^{(+)} ({\bf r}, t) |0 \rangle  = \rho ({\bf r}, t) \, \delta,
\ee
with $\rho ({\bf r}, t)$ the charge density  and $\delta$ the (average) dipole length. By writing the NG condensate density as $\langle 0| \widehat{P}^{\dag} ({\bf r}, t)  \widehat{P} ({\bf r}, t) |0 \rangle = n ({\bf r}, t) $, we have $\rho ({\bf r}, t) \, \delta = |v ({\bf r}, t)|^2 = 2{\cal P} ({\bf r}, t) \, n ({\bf r}, t)$.

We then write the charge density wave function $\sigma ({\bf r}, t)$ as
\be\lab{vs9a}
\sigma ({\bf r}, t) = \sqrt{\rho ({\bf r}, t)}\, e^{i\theta({\bf r}, t)}  ~,
\ee
with real $\rho ({\bf r}, t)$ and $\theta({\bf r}, t)$. We have $ |\sigma ({\bf r}, t) |^2 = \rho ({\bf r}, t) \equiv N({\bf r}, t)$  in the system ground state. Thus,  $N({\bf r}, t)$ denotes the charge density condensate.  Sometimes $\sigma ({\bf r}, t)$ is also called the {\it macroscopic wave function}. We observe that the order parameter $v ({\bf r}, t)$ provides a measure of the NG mode density $n ({\bf r}, t)$ in the ground state.

In terms of $\sigma $, the  local gauge transformation (\ref{2.7}) is
\be  \label{g3.lp2}  \sigma ({\bf r}, t) \,\rar\, e^{- i q\, \la_3 ({\bf r}, t)} \sigma
({\bf r}, t)~,
\ee
Note that the transformation (\ref{2.7}) is induced by the phase transformation
\be\lab{2.20}
\theta ({\bf r}, t) \rightarrow \theta ({\bf r}, t) - q\, \la_3  ({\bf r}, t) ,
\ee
when Eq.~(\ref{vs9a}) is used. One may show \cite{DelGiudice:1985} that the phase $\theta({\bf r}, t)$ represents the NG wave dipole field, also called the phason field \cite{Umezawa:1993yq,Leplae}.

As mentioned, the meaning of the macroscopic wave function, or the order parameter $v ({\bf r}, t)$, resides in the fact that in the presence of the spontaneous breakdown of symmetry, the microscopic quantum components of the system behave in collective or {\it coherent} way, namely they undergo {\it in phase} motion, so that the system is then  globally described by such a common phase of its quantum constituents. This establishes the link between the system macroscopic wave function $\sigma ({\bf r}, t)$ and the microscopic one, $\psi ({\bf r}, t) = \sqrt{a ({\bf r}, t)} \exp(i {S}/{\hbar})$, for the quantum components.  We assume that  $\psi ({\bf r}, t)$ satisfies  the Schr\"odinger equation.  Thus we will set the phase $\theta({\bf r}, t) \equiv {S}/{\hbar}$\,. Then for the momentum ${\rm\bf P} = m {\rm\bf v}$ we have ${\rm\bf P} \equiv {\mbox{\boldmath $\nabla$} S} = \hbar {\mbox{\boldmath $\nabla$}} \theta $. In the presence of the ${\rm\bf A}$ field, the canonical momentum is known to be $m {\rm\bf v} = {\rm\bf P} - q {\rm\bf A}$ (the minimal coupling), thus (cf. Eq.~(\ref{v}))
\begin{eqnarray}\lab{P}
	- \frac{m}{q} {\rm\bf v} =  {\rm\bf A} - \frac{\hbar}{q} {\mbox{\boldmath $\nabla$}} \theta.
\end{eqnarray}
We note that the r.h.s. of Eq.~(\ref{P}) can be considered as a gauge transformation with $\theta ({\bf r}, t)$ being the gauge function: ${\rm\bf A} \rightarrow {\rm\bf A}' = {\rm\bf A} - \frac{\hbar}{q} {\mbox{\boldmath $\nabla$}} \theta$. The requirement that the gauge condition ${\mbox{\boldmath $\nabla$}}\cdot {\rm\bf A} = 0$ be invariant under gauge transformations, i.e. ${\mbox{\boldmath $\nabla$}}\cdot {\rm\bf A}' = 0$,  implies that $\theta ({\bf r}, t)$ (and $\la_3  ({\bf r}, t)$) has to be a solution of the equation ${\mbox{\boldmath $\nabla$}^2} \theta = 0$ (and ${\mbox{\boldmath $\nabla$}^2} \la_3  ({\bf r}, t) = 0$). In such a case,  Eq.~(\ref{P}) gives ${\mbox{\boldmath $\nabla$}}\cdot {\rm\bf v} = 0$.
We remark that in general one may consider the so-called boson transformation
\be\lab{2.20bt}
\theta ({\bf r}, t) \rightarrow \theta ({\bf r}, t) - c\, \ f({\bf r}, t) ,
\ee
with $c$ a convenient constant and $f({\bf r}, t)$, called boson transformation function,  solution of the equation for $\theta$, ${\mbox{\boldmath $\nabla$}^2} f({\bf r}, t) = 0$, so that Eq.~(\ref{2.20bt}) is an invariant transformation for the theory. Such a transformation describes the space-time-dependent (i.e. non-homogeneous) boson transformation and it can be shown \cite{Freeman2010,DifettiBook,Matsumoto:1975fi,Matsumoto:1975rp,Umezawa:1993yq} that in order to have observable effects $f  ({\bf r}, t)$ has to carry a topological singularity. This theoretical frame thus predicts the appearance of topological structure formation, such as vortices, as further discussed in Section \ref{IV} (and in Ref. \cite{Freeman2010}). In the case of the U(1) phase symmetry mentioned above, the
stationary function $f(x)$  may carry indeed a
vortex singularity given by
\be \lab{g3.ts1vo} f(x) = \arctan \left(\frac{x_{2}}{x_{1}}\right)~.
\ee
Eq.  (\ref{g3.ts1vo}) shows that the phase is undefined on the line
$r = 0$, with $r^{2} = x^{2}_{1} + x^{2}_{2}$ (which, as observed in \cite{Freeman2010}, reflects in the
observed phase indeterminacy in the process of transition between
two AM pattern frames).
For an analysis of the vortex
properties associated to Eq.  (\ref{g3.ts1vo}) see Refs.
\cite{Matsumoto:1975fi,Matsumoto:1975rp} and the discussion in Section \ref{IV}.

In the following we study the variation in space and in time of  the macroscopic wave function 
and of the phason field 
condensate. In QFT these variations denote transformations through physically different vacua, i.e. unitarily inequivalent vacua, or, in other words, they denote ``phase transitions". Of course, going through phase transitions, the system moves toward the equilibrium, or stationary regime, which is the one minimizing the free energy density functional $F(\si, \si^{*}, {\rm\bf A} )$  with respect to $\si^{*}$: ${\partial F }/{\partial\sigma^*} = 0$. However, during the phase transition processes such a minimization condition is not satisfied, ${\partial F }/{\partial\sigma^*} \neq 0$. Here we are interested in such  dynamical transition processes since the brain is a far from the equilibrium system,  indeed.

Let us start by considering the Schr\"odinger equation for $\sigma ({\bf r}, t)$:
\begin{eqnarray}\lab{mwf}
	i\hbar\frac{\partial\sigma }{\partial t} = {\hat{H}}\sigma,
\end{eqnarray}
with
\begin{eqnarray}
{\hat{H}} \, \equiv \, \frac{1}{2 m}\left(-i\hbar\mbox{\boldmath$\nabla$}- q {\rm\bf A}\right)^2 + U, \label{Sch.eq}
\end{eqnarray}
where the potential energy $U$ is
\begin{eqnarray}
	U= q\, \varphi+\mu_{0}.
\end{eqnarray}
Here $\varphi \equiv {V}/{q}$, with $V$ the scalar potential and $\mu_{0}$ the chemical potential. $\mu_{0}$ is included in the Schr\"odinger's equation for  $\sigma$ in order to account for the variations in the number $|\sigma|^{2} = N$. In the following our discussion we will closely follow Ref. \cite{Barybin}, where the time-dependent Ginzburg-Landau equation for the modulus of the order parameter is obtained and  conservation laws as well as dissipative effects are taken into account.

The order parameter and the macroscopic wave function $\sigma$ are
not affected by quantum fluctuations (in this sense they are {\it macroscopic} fields). However, as said above,  we are interested in their variations occurring in the (non-equilibrium) phase transition processes. Therefore, we need to allow variations in the number $N$, which justify the introduction of the chemical potential $\mu_{0}$ in Eq.~(\ref{Sch.eq}). We note that a chemical potential term is not included in the Schr\"odinger's equation for the microscopic wave function $\psi ({\bf r}, t)$ describing the elementary dipole components, whose number is assumed to be constant during the system time evolution.

Use of Eq.~(\ref{vs9a}) into Eq.~(\ref{mwf}) gives the the equations for the imaginary and the real parts:
\begin{eqnarray}\lab{imag}
	\frac{\partial N}{\partial t}+\mbox{\boldmath$\nabla$}\cdot\left(N{\rm\bf v}\right)=0,
\end{eqnarray}
\begin{eqnarray}\lab{real}
	\mu_{0}+\mu_{1}+q\,\varphi+\frac{mv^2}{2}
	+\hbar\frac{\partial\theta}{\partial t}=0,
\end{eqnarray}
respectively.  Here and in the following $v^2 = {\rm\bf v} \cdot {\rm\bf v}$. Eq.~(\ref{imag}) is the continuity equation (to be compared with Eq.~(\ref{cont})).
The additional contribution $\mu_{1}$ to the chemical potential  in Eq.~(\ref{real}) is
\begin{eqnarray}\lab{mu1}
	 \mu_{1}=-\frac{\hbar^2}{2m}\frac{\mbox{\boldmath$\nabla$}^2N^{1/2}}{N^{1/2}}
\end{eqnarray}
and is due to non-homogeneous density of the condensate. From Eq.~(\ref{real}) one obtains (see Appendix \ref{A})
\begin{eqnarray}\lab{Lorentzforce}
	m\frac{d{\rm\bf v} }{dt} \, = \, q\, \left({\rm\bf E}+{\rm\bf v} \times{\rm\bf B}\right)
	-\mbox{\boldmath$\nabla$}\left(\mu_{0}+\mu_{1}\right),
\end{eqnarray}
with
\begin{eqnarray}
	{\rm\bf E}=-\frac{\partial \rm\bf A}{\partial t}-\mbox{\boldmath$\nabla$}\varphi
	\qquad {\rm and}\qquad
	{\rm\bf B}=\mbox{\boldmath$\nabla$}\times{\rm\bf A} = - \frac{m}{q} (\mbox{\boldmath$\nabla$}\times{\rm\bf v}).
\end{eqnarray}

We observe that, provided that ${\mbox{\boldmath $\nabla$}}\cdot {\rm\bf v} = 0$ (see the remarks after Eq.~(\ref{P}) and  Eq.~(\ref{divv})), Eq.~(\ref{imag}) gives
\begin{eqnarray}\lab{imag2}
\frac{d N}{d t} =	\frac{\partial N}{\partial t}+ {\rm\bf v}\cdot \mbox{\boldmath$\nabla$} N = 0,
\end{eqnarray}
which expresses the charge density conservation in the stationary regime. Since $N$ is related to the order parameter, this means that the long range correlation, namely the ordering, is preserved in time. In the frame of the dissipative many-body model, such a case describes stationary neuronal correlates. However, observations of the functional activity of the brain show that a succession of neuronal correlates (``wave packets''), modulated in amplitudes and in phase, occurs. In its functional activity the brain appears undergoing a continuous stream of phase transitions. We are therefore interested in studying the non-stationary dissipative phase transition processes, where, contrarily to Eq.~(\ref{imag2}), ${d N}/{d t} \neq 0$. Thus we need to study the non-stationary or time-dependent Ginzburg-Landau equation in order to consider the possibility of dissipative processes.

\section{Time-dependent Ginzburg-Landau equation}\lab{III}

Let us start by considering the Ginzburg-Landau (GL) functional $F (\sigma, \sigma^{*}, \rm\bf A)$ representing the free energy density. For our task we do not need to specify the explicit form of $F (\sigma, \sigma^{*}, \rm\bf A)$. This will depend on the particular dynamical model one adopts for the description of the system under study. In general, it is a non-linear functional of the fields, containing a kinetic energy term $(-i\hbar\mbox{\boldmath$\nabla$}-q {\rm\bf A})^2/2m$ and some potential term.
The stationary Ginzburg-Landau (GL) equation is obtained by extremizing the free energy density $F (\sigma, \sigma^{*}, \rm\bf A)$ with respect to $\si^{*}$:
\begin{eqnarray} \lab{SGL1}
	\frac{\partial F }{\partial\sigma^*} = 0 ,
\end{eqnarray}
where, in full generality, we may write:
\begin{eqnarray} \lab{SGL}
	\frac{\partial F }{\partial\sigma^*}
\, \equiv \,
	\left[\frac{1}{2 m }(-i\hbar\mbox{\boldmath$\nabla$}-q {\rm\bf A})^2
	+\alpha+\beta|\sigma|^2\right]\sigma,
\end{eqnarray}
with $\alpha(T)$ and $\beta(T) = {|\alpha(T)|}/{N^{\circ}(T)} > 0$ acting as a mass term and a positive coupling constant term, respectively. $N^{\circ}(T) = |\sigma|^2$ is the equilibrium condensate density. Their values depend on the temperature $T$ and their explicit expressions depend on the system under consideration and on the adopted phenomenological model (in solid state physics $|\alpha(T)|$ is related to the chemical potential $\mu_0$, expressing, e.g. in superconductivity \cite{Barybin}, the vacuum permeability, and $\beta(T)$ to the critical (magnetic) fields acting upon the system.  Notice that the
contribution $(\alpha+\beta|\sigma|^2)\sigma$ in the r.h.s. of Eq.~(\ref{SGL}) may be considered as derived from the potential $V(\sigma, \sigma^*) = - (\beta/2)(|\sigma|^2 + \al/\beta )^2$. The (mean value of the) "order parameter" $\sigma$ minimizing the potential $V(\sigma, \sigma^*)$ is zero (disordered or symmetric ground state) for $\al \ge 0$, and non-zero for $\al < 0$, with $|\sigma|^2 = - \al/\beta \neq 0$ (the ordered or asymmetric ground state)\footnote{We also note the strong analogy with
the laser phase transition \cite{Haken:1984a}, where the potential $V(\sigma, \sigma^*)$ is again considered. In the phase transition between
the disordered and the ordered (laser) state (the lasering process), the order parameter
$|\sigma|^2$ changes in time from zero to a non-vanishing value proportional to
$\al$, going through the threshold set at $\al = 0$. In the Haken interpretation, $\al$ is the pump parameter
whose tuning may carry the system in the lasering region; $\sigma$ denotes the amplitude of the classical electromagnetic mode.}

Eq.~(\ref{SGL1}) expresses the condition for the stationary (equilibrium) dynamical regime, where the evolution of  the macroscopic wave function is controlled  by the Sch\"odinger equation (\ref{mwf}). Instead, in the case of non-stationary regime, the non-vanishing quantity ${\partial F }/{\partial\sigma^*}$ expresses the rate at which $\sigma ({\bf r}, t)$ approaches its stationary value at the minimum of the free energy. It is then customary to consider the generalized time-dependent GL  (TDGL) equation:
\begin{eqnarray}\lab{SGL12}
	{i\hbar} \frac{\partial \sigma}{\partial t}
= \hat{H}\sigma -\frac{i}{\gamma}\frac{\partial F }{\partial \sigma^{*}} ~,
\end{eqnarray}
where now ${\partial F }/{\partial \sigma^{*}}$ is not zero,  $\ga$ is a relaxation parameter and the stationary regime is reached in the limit $(1/\gamma)\partial F /\partial \sigma^{*} \rightarrow 0$. In such a limit, the equilibrium wave function is $\sigma^{\circ}(t) = |\sigma^{\circ}| \exp(- i\epsilon t/\hbar)$, with $\hat{H}\sigma^{\circ} = \epsilon \sigma^{\circ}$. The second term in the r.h.s. of Eq.~(\ref{SGL12}) thus describes the dissipative contribution coming from incoherent relaxation processes \cite{Barybin}.
Using $\sigma=\sqrt{N}\exp({i\theta})$,  the imaginary and real part of Eq.~(\ref{SGL12}) give   (see the Appendix \ref{B}) the continuity equation:
\begin{eqnarray}\lab{continuity}
	\frac{\partial N }{\partial t}
	+\mbox{\boldmath$\nabla$}\cdot(N {\rm\bf v} )
	=-\frac{2G}{\tau_{\rm GL}}N
\end{eqnarray}
and the relation for the chemical potentials
\begin{eqnarray}\lab{chempot}
	\mu_{0}+\mu_{1}+\mu_{2}+q \varphi+\frac{m v^2 }{2}
	+\hbar\frac{\partial \theta}{\partial t}
	=0,
\end{eqnarray}
respectively, where
\begin{eqnarray}\lab{chempot2}
	\mu_{2}=-\frac{\hbar}{2 \gamma N }
	\mbox{\boldmath$\nabla$}\cdot(N {\rm\bf v} )
	\equiv -\frac{\xi^2_{\rm GL}}{\tau_{\rm GL}}\frac{m}{q}
	\frac{\mbox{\boldmath$\nabla$}\cdot {\rm\bf J} }{N }~,
\end{eqnarray}
is related to the relaxation parameter $\ga$,  is of dissipative nature and proportional to non-homogeneities of the condensate. We have used the definition ${\bf J} \equiv q N {\rm\bf v}$.
In Eq.~(\ref{continuity}) $G$ is given by
\begin{eqnarray}\lab{G}
	G=\xi^2_{\rm GL}\frac{2m}{\hbar^2}\left(\frac{m v^2 }{2}+\mu_{1}\right)
	+\frac{N }{N^{\circ} }-1~.
\end{eqnarray}
and ${\tau_{\rm GL}}$ and $\xi^2_{\rm GL}$ are the GL relaxation time and coherent length, respectively (see Eqs.~(\ref{B1})).
By operating with the gradient operator on the Eq.~(\ref{chempot}) we obtain
\begin{eqnarray}\lab{hydr}
	m\frac{d{\rm\bf v} }{dt}
	=q \left({\rm\bf E}+{\rm \bf v} \times{\rm\bf B}\right)
	-\mbox{\boldmath$\nabla$}(\mu_{0}+ \mu_{1} + \mu_{2}).
\end{eqnarray}
Eqs.~(\ref{continuity}) and (\ref{hydr}) have to be compared with Eqs.~(\ref{imag}) and (\ref{Lorentzforce}), respectively.

By using Eq.~(\ref{G}) and $\mbox{\boldmath$\nabla$}\cdot {\rm\bf v} = 0$, the continuity equation  (\ref{continuity}) can be rewritten as
\begin{eqnarray}\lab{nst}
	\frac{dN }{dt}
	\equiv \left(\frac{\partial}{\partial t}
	+{\rm\bf v} \cdot\mbox{\boldmath$\nabla$}\right)N = - \Gamma_{\rm R}N
	\neq 0
\end{eqnarray}
where we have set
\begin{eqnarray}
	\Gamma_{\rm R}\equiv\frac{1}{\tau_{\rm R}}
	=\frac{2G}{\tau_{\rm GL}}
\end{eqnarray}

We thus see that the relaxation term $\mathcal{R}_{\rm diss} \equiv \Gamma_{\rm R}N$ accounts for the rate of change of the condensate ${dN }/{dt}$.
From the expression of $G$, Eq.~(\ref{G}), we see that we may write  $\Gamma_{\rm R} = \Gamma_1 + \Gamma_2$,  where
\begin{eqnarray}
	\Gamma_1=\frac{\xi^2_{\rm GL}}{\tau_{\rm GL}}\frac{2m}{\hbar^2}mv^2
	=2D_{\rm GL}\left(\frac{m v }{\hbar}\right)^2,
\end{eqnarray}
with the diffusion coefficient $D_{\rm GL} \equiv {\xi^2_{\rm GL}}/{\tau_{\rm GL}}$, and $\Gamma_2$  related with non-homogeneities ($\mu_1$ and $N \neq N^{\circ}$). $\Gamma_2$ is related with dissipative processes whose life-time is usually longer than the GL relaxation time $\tau_{\rm GL}$. In general, the condition $\tau_{\rm R} = \tau_{\rm GL}/2G \gg \tau_{\rm GL}$, requiring, in order to hold, small values of $G$, ensures fast formation of quasi-local equilibrium in the condensate, controlled by $\tau_{\rm GL}$, compared with longer decay time of the condensate density $N$. At a critical  temperature $T_c$, a fast transition to the equilibrium condensate regime occurs; one may then expect
$G \rightarrow 0$,  so that  $\tau_{\rm GL} \ll \tau_{\rm R} = \tau_{\rm GL}/2G$, as $T \rightarrow T_c$.

Eq.~(\ref{nst}) gives the general TDGL equation for the normalized wave function $\chi=|\sigma|/|\sigma^{\circ}|\equiv(N /N^{\circ} )^{1/2}$:
\begin{eqnarray}\lab{nss}
	D_{\rm GL}\mbox{\boldmath$\nabla$}^2\chi-\frac{d\chi}{dt}
	=- \frac{1}{\tau_{\rm GL}}\left[(1-\chi^2)-\xi^2_{\rm GL}\left(\frac{m v }{\hbar}\right)^2\right] \chi~,
\end{eqnarray}
where
$d\chi/dt=\partial\chi/\partial t+{\rm\bf v} \cdot\mbox{\boldmath$\nabla$}\chi$.
In the approximation of $\Ga_{\rm R} \approx  \Ga_{1} \equiv 2/{\tau_{\rm E}}$, 
and in the case of fast formation of quasi-local equilibrium, we have
\begin{eqnarray}\lab{tll}
	\frac{\tau_{\rm GL}}{\tau_{\rm E}}
	=\xi^2_{\rm GL}\left(\frac{m v }{\hbar}\right)^2
	\ll 1.
\end{eqnarray}
In such a limit, for $\tau_{\rm GL}\ll\tau_{\rm E}$,  Eq.~(\ref{nss}) becomes
\begin{eqnarray}
	D_{\rm GL}\mbox{\boldmath$\nabla$}^2\chi-\frac{d\chi}{dt}
	=-(1-\chi^2)\frac{\chi}{\tau_{\rm GL}}~,
\end{eqnarray}
which further reduces to
\begin{eqnarray}\lab{sGL}
	\xi^2_{\rm GL}\mbox{\boldmath$\nabla$}^2\chi+\chi-\chi^3=0
\end{eqnarray}
when the stationary condition $d \chi/d t = 0$ is met, i.e.  $\partial\chi/\partial t=0$ and
${\rm\bf v} \cdot\mbox{\boldmath$\nabla$}\chi=0$. Eq.~(\ref{sGL}) reproduces indeed the stationary GL equation (\ref{chi}) 
when the disequality (\ref{tll}) holds.

\section{The vortex solution}\lab{IV}

Our discussion in the previous Sections concerning the stationary regime and the departure from it occurring in the phase transition processes is centered on the behavior of the quantities $\tau_{\rm GL}(T)$, $\xi^2_{\rm GL}(T)$ and $\eta_{\rm GL}(T)$ defined in Eqs.~(\ref{B1}). These quantities, in turn, depend on the behavior of $|\al (T)|$, which, as we have observed in Section \ref{III},  has the meaning of the  mass term in the equations for the macroscopic wave function (the order parameter) $\sigma$, or the  wave function $\chi$ normalized to $|\sigma^{\circ}|$, namely $\chi=|\sigma|/|\sigma^{\circ}|\equiv(N /N^{\circ} )^{1/2}$  satisfying the general TDGL equation (\ref{nss}) (equivalent to Eq.~(\ref{nst})). We might consider non-instantaneous phase transition processes where a time-dependent temperature might still be defined. In these processes, assume that the transition starts at the critical temperature $T_C$ and the stable configuration is reached at the so-called ``Ginzburg temperature" $T_G$, with  $T_G < T_C$, after a certain interval of time during which the system is said to be in the critical regime \cite{Bettencourt}. 
In practice, the model here adopted aims at describing brain functioning
as a repeated passage through a critical temperature state towards
which the brain naturally relaxes  after having reached a lower-temperature
stable configuration, that corresponds to the response to some external
stimulus or any other kind of perturbation. The crucial point is that this ``critical dynamics" is always performed as a transition between different
thermodynamic equilibrium states (at $T_C$ and at $T_G$): irrespectively of
the nonequilirium processes involved into this transition, this implies
that the general thermodynamic description derived from the
fluctuation theorem \cite{Crooks} applies to this model of critical dynamics
of the brain. The rate of entropy production that is inherent in the process 
described by the TDGL equation can be accordingly related to the amount of
heat dissipated by the brain to reach new equilibrium configurations below
the critical temperature $T_C$.

We want to model time dependence of $|\al (T)|$ during such a critical regime evolution.

Here we remark that the departure from the stationary regime (at $T_C$), representing the system ground state at a given time in the system
evolution\footnote{See Refs. \cite{Vitiello:1995wv,Vitiello:2001,11} for the discussion of the existence of infinitely many vacua (ground states) representing different, physically inequivalent, microscopic configurations of the system.}, namely the start of the critical regime, is driven by fluctuations which may be spontaneous (typically occurring in a quantum vacuum) and which can trigger a phase transition, if the requisite transition energy is provided by some  endogenous or external stimulus. These ground state fluctuations turn into temperature fluctuations since in the dissipative model the ground state is in fact a thermal state \cite{Vitiello:1995wv,Vitiello:2001,11}. At the end of the critical regime (at $T_G$) the systems arrives at a ``new" ground state configuration and the ``phase transition" is thus completed. As a matter of fact, the brain system undergoes a continuous sequence of phase transitions in a path through the (infinitely) many ground states whose existence is allowed by QFT \cite{Vitiello:1995wv,Vitiello:2001,11}.

This is a delicate point, where the dissipative many-body model turns out to be very helpful in depicting what a traditional non-equilibrium  thermodynamical approach cannot explain. It is well known that mammalian brains operate at a steady state constant temperature that is homeostatically regulated. Then it is crucial reconcile the temperature invariance of brains with the
temperature fluctuations intrinsic to ``far-from-equilibrium" activity seen in 
laboratory experiments as well as in everyday life. (As noted in Section I, brains consume free energy at rates ten-fold greater than any other
organ caused by the heat exchanges at a local
molecular level due to the ceaseless
electrochemical and metabolic reactions).
The resolution of the apparent contradiction between homeostasis and the far from the equilibrium brain activity is implicit in the dissipative many-body model, which suggests that a phase transition entails a brief, local fluctuation of temperature during the transition that punctuates the pseudo--equilibrium
steady state. A peak is followed by reversal (a 
so-called biphasic transient), leading the system to a state that differs from the one before, as observed and predicted by the model. (For brevity we do not report more on such a hysteresis-like property in  brain dynamics; see e.g. \cite{Vitiello:2004,Manka:1990,Pessa:2004,Pessa2003,11}).

In fact, such a transient fluctuation in temperature has been demonstrated for nerve and
modified muscle otherwise held in thermal steady state in experiments with the squid giant axon
and siphon \cite{Abbot}. It shows that an action potential is accompanied by  brief cooling of the nerve and synapse when the sodium ions expand into the intracellular compartment and the potassium ions expand outwardly. The cooling is soon overshadowed by the heat released by the burst of metabolism that pumps the ions back to their compartments and restores the energy
reserve. The thermal reservoir of the water provides the buffering and averaging, such that the dissipative system can undergo substantial local fluctuations in temperature, which are masked in the global heat production. Thus we see another crucial role of water by continually holding temperature constant in the average, although locally (in space and in time) fluctuating.  Our reference to the dependence of the system on temperature has to be understood as its dependence on such localized transients.

It is interesting to note that in completely physically different contexts (such as in neuroscience, condensed matter physics, particle physics and cosmology) intensive theoretical and experimental research has shown that extended objects with non trivial topology (as for instance vortices) appear during the critical transition processes (see e.g. \cite{kib2,zurek1,Alfinito:2002a,Alfinito:2002b}) and persist for varying time intervals thereafter.
In the present Section we are indeed going to discuss the vortex solution to the TDGL equation.

Due to the nonlinearity of the TDGL,   numerical simulations  have been associated often to the theoretical analysis (see \cite{Bettencourt} and Refs. quoted therein). Let us write  Eq.~ (\ref{nss}) as
\begin{eqnarray}\lab{nssv}
\frac{1}{D_{\rm GL}}\frac{d\chi}{dt} =
	\mbox{\boldmath$\nabla$}^2\chi + \frac{1}{\xi^2_{\rm GL}}\left[(1-\chi^2)-\xi^2_{\rm GL}\left(\frac{m v }{\hbar}\right)^2\right] \chi~,
\end{eqnarray}
which admits as a solution the (quasi-)static vortex solution for $\frac{1}{D_{\rm GL}}\frac{d\chi}{dt} = \hbar \ga \frac{d\chi}{dt} \approx 0$ (in such a case, indeed, it is recognized to be the vortex equation) \cite{Manka:1990,DifettiBook}. The condensate produces the classical vortex envelope $\sigma (r)$. Since $\sigma_{0} $ is temperature dependent, the vortex envelope changes with $T$. The phase transition and the vortex formation appear thus as the effect of localized (non-homogeneous) boson condensation.

Actually, the phase transition begins with an abrupt, adiabatic, localized decrease of the order parameter (a measure of the analytic power of the background activity) to near zero, denoted as the {\it null spike}, resulting locally in loss of order and coherence, hence onset of symmetry. Concomitant to this, the spatial variance of the analytic phase increases and a discontinuity in space and in time of the analytic phase appears. The start of the phase transition is thus adiabatic and it appears to be instantaneous at the time scale of the beta and gamma cycle durations.

The extreme spatiotemporal localization of the null spikes indicates that they are associated to singularities. These constitute the core out of which, in the non-instantaneous phase transition process,  the vortices start to form. The singularities coincide with the apex of the phase cones which start to develop and spread over the system. We can thus say that the null spikes mediate or precipitate the phase transition toward the formation of a new AM pattern. The non-homogeneous boson condensation at the origin of the formation of vortices and ordered localized patterns requires energy (ordered states are lower in energy and separated by an energy gap from the symmetric state). Consequently, apart the adiabatic instantaneous start, the
phase transition itself appears to begin with cooling and is non-instantaneous. The cooling appears as a manifestation of the process of symmetry breakdown eventually leading to the emergence of new extended AM and PM patterns. It is thus different from the very localized cooling that may occur due to localized (e.g. on the axon) metabolic activity.

It is interesting to observe that these features predicted by the model  find a correspondence in the features observed at the level of neuronal assemblies. It appears  that the neural transition requires three conditions as precursors. The first is macroscopic arousal with a high level of
background 1/f activity (see our remarks in the following Section). Second is activation of a Hebbian nerve cell assembly, which amplifies,
generalizes and abstracts a sensory signal into a
burst of high intensity but microscopic firing of
pulses. Third is the initiation of a narrow-band
oscillation in that burst. These three changes
require an increase in dissipation of free energy
as heat and therefore in temperature. But the
phase transition itself begins with cooling as predicted by the dissipative model.

From the TDGL equation (see also Ref. \cite{Freeman2010} and \cite{Manka:1990}) one may also derive that, in the first approximation the vortex core size is of the order $\xi^2_{\rm GL} \propto |\al|^{-1}$. This means that the vortex core is increasing with temperature increase, as $T$ approaches  $T_{C}$ from below, the vortex envelope disappearing at $T_{C}$. Rising temperature above $T_{C}$ leads to symmetry restoration (unstructured ground state).

Vice-versa, going from above back to $T_{C}$ the unstructured background activity (fully symmetric) with vanishing or very low analytic amplitude exhibits, at $T_{C}$, undefined analytic phase, namely the singularity (null spike) as it is at the center line of the vortex core. As $T$ is lowered below $T_{C}$, the critical regime starts, vortices appear, whose core shrinks as temperature further decreases.

As mentioned in Section II the singular NG boson (the phason) condensation is essential for obtaining a nontrivial topological charge. In Refs. \cite{Matsumoto:1975fi,Matsumoto:1975rp} it is shown that the
space-time-dependent order parameter can be gauged away by an appropriate gauge
transformation when the boson condensation function is regular. When the boson condensation function presents singularities corresponding to regions occupied by the normal state (i.e. without condensation) the boson condensation is manifested in macroscopic structures. The vortex, e.g., is singular on the z-axis, at $r = 0$. In this case, the topological charge is characterized by the integer "winding numbers" $n \neq 0$.  Associated to the vortex there is the quantized flux $\phi = 2\pi \,n/q$ \cite{Manka:1990,DifettiBook,Freeman2010}.

When the l.h.s of Eq.~(\ref{nssv}) is not neglected, one can show that in order to have a solution behaving as $\sigma \propto \exp -(\om/\hbar\ga)t$, the condition
\be {\bf k}^{2} \ge m^{2}(t)~, \lab{g5.21} \ee
has to hold during the critical regime between $T_C$ and $T_G$ for each $k$-mode ($k \equiv \sqrt {{\bf k}^{2}}$), with ${\bf k}$ the wave number and $m^{2}(t)$ the time-dependent ``effective mass". In our discussion then we may follow the arguments presented in Ref. \cite{Freeman2010}, where we have considered the formation of vortices during the critical regime and the appearance of null spikes (there we have considered the harmonic limit approximation consisting in neglecting the nonlinear term in the TDGL equation (\ref{nss}) \cite{Bettencourt,Alfinito:2002a,Alfinito:2002b}).

Suppose the
critical regime starts and ends at the times $t=0$ and $t= \tau$, respectively. For a given ${\bf k}$, Eq.~(\ref{g5.21}) holds up to a time $\tau_{k}$ such that  $m^{2}(t)$ is larger than ${\bf k}^2$ for $t > \tau_{k}$. The
corresponding $k$-mode can propagate in a span of time $0\le\ t \
\le \tau_k$, which determines
the dimensions to which the domains can expand and the ``effective causal horizon" \cite{kib2,zurek1}
will be inside the system (possible formation of more than
a domain) or outside (single domain formation) according to whether
the time occurring to the $k$-mode to reach the boundaries of the system is
longer or shorter than the allowed propagation time, respectively.

    In order to determine the value of $\tau_k$, one must assign the explicit form of
$m^2 (t)$. Its time-dependence may be fixed in such a way to allow  vortex
formation \cite{Alfinito:2002b}. One possible modeling of $m^2 (t)$ is then:
\be  m^{2}(t)  = m_{0}^{2} \ e^{2h(t)} ~. \lab{g5.151} \ee

Let $\xi$ denote the correlation
length corresponding to the $k$-mode propagation and $L \propto
m_{0}^{-1}$. Then the correlation propagation time is implicitly obtained from Eqs.~(\ref{g5.21})  and (\ref{g5.151}) as:
\be h(\tau_k) =  \ln\left(k  \over m_{0} \right) \; \propto  \;
\ln\left( L\over \xi \right)~. \lab{g5.221} \ee
We see that $L$ acts as an intrinsic infrared cut-off. Small
values of $k$ are indeed excluded by the non-zero minimum value $m_{0}$ of $m$.  Consequently, long wave-lengths
are not allowed. This means that only domains of finite size can be obtained, which is what happens in the brain activity. The linear size of the domains over which the correlation length extends at the end of the critical regime is of the order of $\lambda_{k} \propto m^{-1}(\tau_k)$.

It remains to choose the explicit analytic expression for $h(t)$ in Eq.~(\ref{g5.151}). The choice we may adopt for $h(t)$ is \cite{Alfinito:2002b}:
\be h(t)= \pm \frac{a t}{b t^{2} + c}~, \lab{g5.24a} \ee
where $a, b, c$ are (positive) parameters chosen so to guarantee dimensionless $h(t)$.
Let $\lambda$ be an arbitrary constant and define the
ratios   $c/a\lambda \equiv  \tau_{Q}$, $a\lambda/b \equiv
\tau_{0}$. We have
$h(\tau_{Q}) = h(\tau_{0})$ and the time derivative of $h(t)$ (and thus
of $m^{2}(t)$) is zero at $t = \tau = \pm \sqrt{\tau_{Q}\tau_{0}}$, which thus plays the role of the equilibrium time scale. We have
\be
h(t) = \pm  \frac{1}{\la \tau_{Q}} \frac{1}{1 +
\frac{t^{2}}{\tau^{2}}} ~t \approx \pm \frac{\Ga}{2} ~t~,
\lab{g5.24} \ee
for $t^{2}/\tau^{2} \approx 1$ and $\Ga \equiv 1/\la \tau_{Q}$.
In the linear approximation one finds that the number of vortices $n_{def}$  possibly
appearing during the critical regime is given  by \cite{Alfinito:2002b,zurek1}:
\be n_{def} \propto m^{2}(\tau) 
\approx m_{0}^{2}\ | \tau/{\lambda \tau_{Q}}|~.
\lab{g5.24ter} \ee

Since the size of the vortex core is given by
$(m(t))^{-1}$, Eqs.~(\ref{g5.151}) and (\ref{g5.24}) show
that such a size evolves in time as $e^{\mp \Ga ~t}$, $t < \tau$ ($t
< \tau_{k}$ for the $k$-mode). Thus, as observed in Ref. \cite{Freeman2010}, we have both,
converging (imploding) and diverging (exploding) regimes, which is in agreement with the phase cone behaviors observed in laboratory.  The ``normal" (disordered) state is confined to the vortex core, then in the case of enlargement of the vortex core
(exploding regime) disorder (local correlations) prevail. Instead, the shrinking of
such a region (imploding regime) may signal that ordering (long range
correlation) is prevailing (the vortex is ``squeezed
out"). This is in agreement with
laboratory observations which show that  explosion or implosion is obtained if the local connections or the long axon connections predominate, respectively
\cite{38}. Observations also show that many phase cones exhibit little or no
rotation but repetitive outward or inward pulsations with each
cycle. When rotational gradients (vortices) are observed, the
singularity is associated to the vortex core singularity.  All four types of these observed spatiotemporal phase gradients (See Fig. 2) are thus predicted by the model.

Finally, we note that conventional neurodynamics may explain the negative gradient  (e.g. in terms of a pacemaker), but not the positive
gradient. Moreover, in the conventional framework  there is no explanation of why both gradients, the positive and the negative one, occur, one or the other at random. Furthermore, we offer as a prediction the expectation that the locations of the initiating null spike, the apex of the following phase cone, and the center of rotation or expansion-contraction of the vortices will be closely related though not necessarily coincident. At present the available experimental techniques indicate that this occurs, but they lack sufficient resolution and precision to provide experimental proof.

\section{Power laws and coherent states}\lab{V}

In this Section we consider one more of the predictions of the many-body dissipative model which is confirmed in laboratory observations. According to the model, the system ground state and the phase transition processes are characterized by the boson transformations, such as the one of the NG field described by Eq.~(\ref{2.20bt})  (see also Eq.~(\ref{2.20})). Boson transformations are the ones by which  (Glauber) coherent states are constructed \cite{Perelomov:1986tf,DifettiBook}. Coherence is therefore a distinctive feature of the basic dynamics of the system and one may expect that it manifests itself at a mesoscopic and macroscopic level since coherent states are known to be the quantum states most near to classical (macroscopic) states. Elsewhere \cite{11a,Freeman2010,11} it has been stressed that one of the merits of the dissipative many-body model is that ``classicality" emerges naturally out of the basic quantum dynamics, not as an artificial classical limit imposed by hand. The laboratory observations are in fact at a classical macroscopic level, as it should be (it has to be stressed that in our analysis neurons, glia cell and other biological structures are classical objects; the quantum variables are the dipole vibrational modes). However, some of the system features are {\it macroscopic quantum features}, in the sense that they cannot be explained without recurs to the underlying quantum dynamics. Among such features, there are the power-laws observed in the brain functional activity, on which  we focus our attention in this Section.

In the frame of the entire analytical function formalism, power-laws have been shown to be related to coherent states \cite{FractalsNMNC,FractalQI}. Let us very briefly summarize such a result. Power-laws denote scale free, self-similar phenomena of fractal nature characterized by a straight line graph in log-log coordinates, with the fractal dimension $d$ corresponding to the slope of the line (the tangent $y/x$ of the straight line angular coefficient). Self-similarity is a characterizing fractal property \cite{Peitgen,BakinBunde} expressed by
\be \lab{selfsim}  (q\,\ga)^n \,=\, 1, \qquad {\rm for~ any~ integer} ~n,
\ee
where for convenience (see below) the number $q$ is parameterized as $q = 1/f^{d}$ and $\ga$ is the parameter characterizing the specific considered fractal, whose dimension $d$ is then given by
\be \lab{d} d ~=~ - \frac{Log \ga}{Log f}~.
\ee
This equation clearly describes a straight line in the log-log coordinates ($x = Log f$, $y = Log \ga$). Here we denote $log_{10}$ by the symbol $Log$. We now remark that $(q\,\ga)^n$, up to the normalization factor $1/\sqrt{n!}$, is the $n$th element of the basis of the entire analytical functions defined in the space of the complex numbers $z = q\,\ga$. In Eq.~(\ref{selfsim}) the number $q\,\ga$ is considered to belong to the restriction to the real axis of the $z$-complex plane. The connection with coherent states is then established by observing that they are in fact constructed in the Fock-Bargmann representation (FBR) by use of the entire analytical functions \cite{Perelomov:1986tf,DifettiBook}. More precisely, the coherent states for such a $z = q\,\ga$ variable are defined to be ``$q$-deformed" coherent states of coherence strength $\ga$, or, also, squeezed coherent states, $q$ being the deformation or squeezing parameter.
We will not insist on such technical points here and the reader may consult Refs. \cite{FractalsNMNC,FractalQI} for details. The point is that, as a result, self-similar processes of fractal dimension $d$ can be  associated to $q$-deformed quantum coherent states, namely they appear as macroscopic systems generated by local deformations of coherent quantum dynamics, they are {\it macroscopic quantum processes} induced by  quantum boson transformations.

\begin{figure}
\centering \resizebox{12cm}{!}{\includegraphics{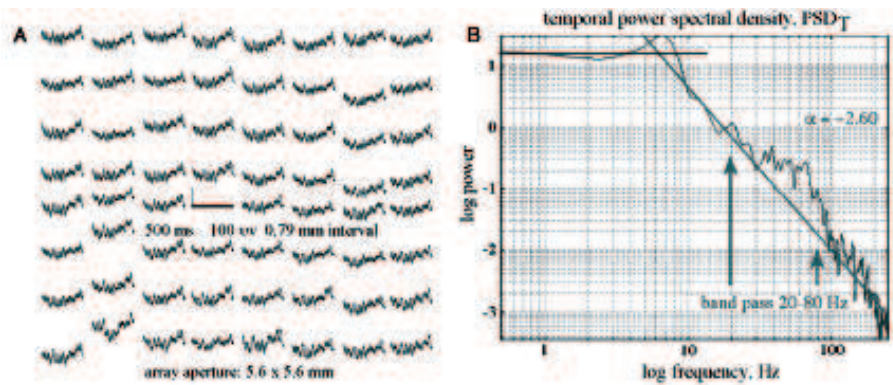}}
\caption{\small \noindent
A. Multichannel recording from a high-density array fixed on the auditory cortex of a rabbit at rest. B. The power spectral densities in the working ECoG were computed for all 64 signals and averaged. The power-law trend lines ($1/f^0$ and $1/f^{2.6}$) were drawn by hand to emphasize the multiple peaks of power in the theta and beta-gamma ranges above the line, which were missing in the resting ECoG. From Figs. A1.01 and A1.02 in \cite{13}} \label{Fig: 2}
\end{figure}

\begin{figure}
\centering \resizebox{12cm}{!}{\includegraphics{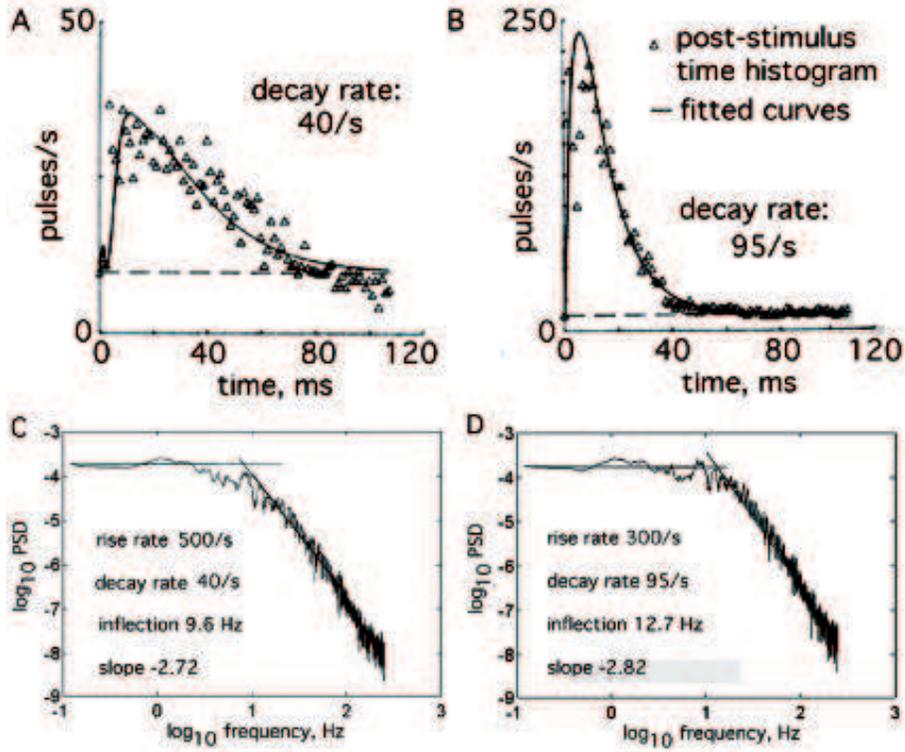}}
\caption{\small \noindent
A, B. The symbols $\Delta$ show the post stimulus time histogram (PSTH) of a representative neuron in response to a weak electric shock, fitted with the solution to a $4$th order linear differential equation \cite{Freeman1975}. The prolonged discharge without inhibitory overshoot is reverberation that is due to mutual excitation. C, D. The decay rate determines the inflection frequency of the power-law $PSD_{T}$, and the rise rate determines the exponent, $\al$. The predicted range of the trend lines is $2 <  \al$ in $1/f^{\al}$. We propose that the segments give the canonical form of resting cortical $PSD_{T}$. From Figs. 5, 7 in \cite{Freeman14}} \label{Fig: 3}
\end{figure}

Let us now turn to the laboratory observations (see e.g. \cite{Freeman1975,9,10,12,13}) and to the graphs of the power laws depicted in the lo-log coordinates in Figure \ref{Fig: 2} and
\ref{Fig: 3}. On the basis of what said above, it is evident that we face macroscopic quantum processes characterized by a coherent dynamics. Coherence is in fact the property to which the analysis in terms of neuronal correlates also leads. This can be seen as follows. In laboratory observations \cite{Freeman1975,9,10,12,13} multichannel recording yields a collection of state variables. Weakly coherent signals are obtained in ECoG with electrodes which are sufficiently far apart. The cortex may be considered to be modular, and each module gives a signal. Coherent signals are instead obtained when the electrodes are closely spaced. In such a case we may talk of variables defining a state space. They are all from the same module, and their variations in time define a trajectory through the modular state space. Many data have been collected \cite{9,10,12,13} on ECoG spatial imaging coming from small, high-density electrode arrays  fixed on the surfaces of the olfactory, visual, auditory, or somatic modules (typically $8 \times 8$ electrodes, spaced $0.79~ mm$ apart, giving a $5.6 \times 5.6~ mm$ square aperture onto the ECoG). Figure \ref{Fig: 2}.A shows the unfiltered background ECoG, obtained in such a way, which is featureless in appearance. Spectral analysis of the ECoG shows broad distribution of the frequency components in Figure \ref{Fig: 2}.B, where the temporal power spectral density ($PSD_T$) in log-log coordinates is power-law, $1/f^\al$, in segments. Below an inflection in the theta-alpha range the $PSD_T$ is flat, $\al = 0$. Above, the $log_{10}$ power decreases linearly with increasing $log_{10}$ frequency  in the beta-gamma range ($12.5-80~ Hz$) with the exponent $\al$ between $2$ and $3$. One can show that in slow wave sleep the exponent averages near $3$ \cite{Freeman14}; in seizures it averages near $4$. Above $75~ Hz$ the slope either increases ($\al=4$, \cite{Miller}), or it flattens to $\al=0$. On the basis of what said above, such values of the slope $\al$ provide corresponding values of the fractal dimension $d \equiv \al$, with deformation parameter $q \equiv 1/f^\al$ and coherent strength $\ga$ corresponding to the power spectral density.
In Figure \ref{Fig: 2}.B, each peak  of power above the 1/f trend line in the beta-gamma range (arrows in Figure  \ref{Fig: 2}.B) reflects a brief epoch of narrow band oscillation. These multiple peaks indicate a departure from the scale free regime (the straight line) and therefore the presence of structures emergent from the background activity.
Figure \ref{Fig: 3}.A, B shows the post stimulus time histograms (PSTH) of the microscopic pulses from a representative neuron in an excitatory population due to low and high intensity electric shocks. The averaging gives the impulse response of the macroscopic population. It presents a rapid rise and an exponential decay to the background firing rate. When the shock strength is reduced to threshold, the decay rate approaches zero. In dynamical terms the asymptotic convergence of activity to the background after perturbation is evidence that a point attractor governs the cortical background activity at unity gain (pp. 285-305 in \cite{Freeman1975}). An attractor by definition represents a stable (ground) state. The power law in Figure \ref{Fig: 3}.C, D signals the scale free (``spatial similarity") property of such ground states. The spatial similarities reveal the long spatial distances across which synaptic interactions can sustain coherence, namely the high-density coordination of neuron firing by synaptic interaction, in agreement with the dissipative model prediction. In the frame of our discussion of theoretical and observational results, the coherent mutual excitation among pyramidal cells thus appears  as an especially attractive conclusion, also considering that they comprise about $80\%$ of cortical neurons, and they provide $90\%$ of the synapses of cortical origin to each other.

\section{Conclusions}\lab{VI}

The brain is an open system submitted to continuous exchange of energy and information with the environment. Sensory inputs drive the brain out of its background activity and large amplitude modulated patterns appear \cite{9,10,12,13} which are the expression of long range neuronal correlation. Moreover the brain manifests a form of conditional stability, criticality, denoting a state of readiness to transit from one state to another, like the one of a subject who is expecting one of several CS, by which to choose one of several courses of action. In the expectant state of searching, a sensory cortex can be viewed as having a set of attractors, one for each expected CS, and each with its basin around an attractor in an attractor landscape \cite{Skarda}. The CS determines the choice by directing the state trajectory into a particular basin. Convergence to the attractor directs the cortex to form the AM pattern that is selected by the CS. The cortical assemblies of neurons can complete widespread phase transitions in very few milliseconds, regardless of their correlation lengths and carrier frequencies. This property can help explain the rapidity of the flow of mental images. The distances across which coherent oscillations are maintained (the correlation lengths) are far larger than the modal diameters of the dendrites of the participating neurons and can include multiple cortical areas, even the entire scalp, which can help to explain the multisensory integration that is required for Gestalt formation. Moreover, spectral analysis of the ECoG shows the occurrence of $1/f^{\al}$ power-law form of the temporal spectral density in log-log coordinates which indicates that the activity is scale-free.

The above properties are the macroscopic manifestation of the mesoscopic and microscopic dynamics. The formation of AM and PM assemblies of neurons have been associated, in previous works \cite{11a,11}, with the spontaneous breakdown of symmetry occurring at the level of the many-body system components and their coexistence and temporal succession to the dissipative character of the dynamics. The large distances across which coherent oscillations are observed to occur in the brain is accounted by the long range of the correlation, extending even to the whole system volume, predicted as a consequence of the spontaneous breakdown of symmetry in QFT. In this paper, continuing our study of the many-body dynamics underlying these behaviors,  we have focused our attention on the non-equilibrium non-instantaneous phase transitions occurring at the level of the many-body dynamics. In the stationary regime (Section \ref{II}), as customary, one considers the extremizing condition of the system free energy $ \frac{\partial F }{\partial\sigma^*} = 0$  whose solutions describe the ground state of the system. In the brain activity, however, the brain is continuously moved out from its  ground state activity entering a non-stationary dynamical regime. We have thus considered (Section \ref{III}) a non-vanishing value for the quantity $\frac{\partial F }{\partial\sigma^*}$ , which now expresses the rate at which the system approaches to the stationary regime. In such a way we have obtained the time-dependent Ginzburg-Landau equation. During such a non-equilibrium phase transition, named the critical Ginzburg-Landau regime, the system dynamics turns out to be characterized by topologically non-trivial structures (vortices), which in the formalism of the dissipative model of brain are described by non-homogeneous boson condensation processes. The size and the life-time of these topological structures have been discussed (Section \ref{IV}) and their behavior commented upon in relation to the temperature changes in the domain of the critical regime, from the critical temperature $T_C$ at which the transition starts to the so-called Ginzburg temperature $T_G$, with $T_G < T_C$. The number of vortices appearing in the phase transition process has been estimated, in agreement with a previous analysis \cite{Freeman2010}. In such a critical phase transition the system moves towards a new stationary ground state, which thus describes an attractor for the system non-linear evolution. Since our frame is the one of QFT, which posses infinitely many unitarily inequivalent ground states, the critical system transition to the stationary regime is not at all described by a trivially determined trajectory in the state space of the system. Elsewhere \cite{11a,11,Vitiello:2004,Pessa2003,Pessa:2004} we have shown that such a trajectory is indeed a deterministic chaotic trajectory in an attractor landscape, the initial conditions being determined by the CS specificity.

Finally, the observation of the  $1/f^{\al}$ power-law of the temporal spectral density in the lo-log coordinate plot, which exhibits a scale free activity in the brain ground state activity, has been described, in agreement to previous studies of the relation between fractals and coherent states \cite{FractalsNMNC,FractalQI}, as the macroscopic manifestation of the underlying coherent many-body dynamics, namely in terms of the scale-free property of coherent states. In such a derivation, the properties of the entire analytical functions have been used.

\appendix

\section{}\lab{A0}

The ECoG and EEG present nearly random but bounded amplitude and spectral distributions, often power-law ($1/f$). The oscillations in the beta-gamma range occur in brief epochs with spatial patterns of amplitude and phase that are statistically constant. In those stationary epochs, the two major operations of normal cortical dynamics of dendritic integration of waves and axonal transmission of pulses are executed in near-linear domains \cite{Freeman1975}. Superposition justifies our use of the tools of linear analysis.  The interactions among excitatory neurons that produce the random $1/f$ oscillations in the ECoG and EEG background activity at rest has been modeled with a single positive feedback loop \cite{Freeman14}, and  the interactions among excitatory and inhibitory neurons that produce the beta-gamma oscillations ($12-80 ~Hz$) with a negative feedback loop \cite{Freeman1975,FreemanErwin}.

Multiple AM/PM patterns coexist in different frequency bands, overlapping in space and time. Whether they interact and if so how strongly has not been determined; they appear to have something like the imperturbability of solitons. The stability of the patterns embedded in and arising from the background turbulence leads us to view cortical dynamics as bistable. In the behavioral state of expectancy the cortex is in a receiving mode that is characterized by disorganized, random pulse firing and chaotic wave activity.
We adopt the AM pattern feature vector as an order parameter. For such a state it is zero, and say that the cortical populations are in a gas-like phase. In the behavioral state of perceiving the cortex is in a transmitting mode of strong interaction, which imposes order on the firing of pulses and synchronization on the carrier wave. The order parameter is non-vanishing positive. The change resembles the phase transition from a gas-like phase to a liquid-like phase of condensation. Usually on close examination of the analytic phase we can find a discontinuity in the time series at which the new AM/PM patterns begin \cite{Freeman2008} suggesting a Type I phase transition. After $\sim 3-5$ cycles of the carrier the patterns more slowly dissipate, suggesting a Type II phase transition.

The transitions only occur when a CS is presented and the subject is in an expectant state. We believe that the phase transition requires a non-vanishing activation energy, strong enough in order to minimize the risk of false alarms and to conserve energy, and that this energy is provided by a Hebbian assembly, which is an intracortical mesoscopic network that has been formed by prior reinforcement learning each discriminable CS. Its functions are to amplify, abstract and generalize each weak sensory input by reverberation (mutual excitation) to activate the entire assembly by any part. The Hebbian assembly also greatly facilitates narrow band oscillation in the beta-gamma range \cite{Freeman1975}. The bandwidth is not zero; the narrow distribution about the center carrier frequency causes beats, which appear as reductions in amplitude that recur at intervals proportional to the bandwidth regardless of the center frequency \cite{Rice1950}. During a beat the power in the designated frequency band can briefly go to zero \cite{FreemanCognitive} in one or more null spikes \cite{Freeman2010}. We propose that the null spike manifests a space-time singularity at which the phase transition begins. Its location may weakly correspond to those of the apex of the phase cone and the center of vortex rotation.

Evidence for the vector fields is derived from recording and interpreting the highly textured spatial patterns of oscillations provided by the images of scalar fields of potential from the EEG and ECoG. The neural mechanisms of vector field formation are modeled with large networks of differential equations that we solve by piece-wise linear approximation (Chapter Six in \cite{Freeman1975} and by implementation in VLSI \cite{Principe}. The solutions show the existence of a point attractor that homeostatically regulates the background activity that maintains cortex in a stable state of criticality. This is the receiving phase \cite{Freeman14}, in which the oscillatory impulse response has en exponential envelope with an exponent, $a < 1$. At criticality $a = 0$, endowing cortex with exquisite sensitivity to selected input, form scale-free, power-law distributions of its parameters and state variables, and establish remarkably long correlation distances that can include the entire cortex of each hemisphere.  In the transmitting state $a > 0$, and the cortex transits to a limit cycle attractor in a form of subcritical Hopf bifurcation. It is the amplified output of the Hebbian assembly that selects the carrier frequency, and it is the null spike that drives the cortex inexorably to the limit cycle attractor.

\section{}\lab{A}

Let $\psi ({\bf r}, t) = \sqrt{a ({\bf r}, t)} \exp(i {S}/{\hbar})$ denote the wave function for the system quantum components, i.e. the {\it microscopic wave function}. It satisfies the Schr\"odinger equation from which the continuity equation
\begin{eqnarray}\lab{cont}
	\frac{\partial a}{\partial t}+{\mbox{\boldmath $\nabla$}}\cdot (a \,{\rm\bf v})=0
\end{eqnarray}
can be derived.
As well known, by defining
\begin{eqnarray}\lab{vQ}
	{\rm\bf v}=\frac{1}{m}\mbox{\boldmath $\nabla$}S \qquad
	{\rm and} \qquad
	Q=-\frac{\hbar^2}{2m}\frac{\mbox{\boldmath $\nabla$}^2 a^{1/2}}{a^{1/2}}~,
\end{eqnarray}
one obtains the hydrodynamic equation of motion (the Madelung picture)
\begin{eqnarray}
	\frac{\partial\rm{\bf v}}{\partial t}+
	({\rm\bf v}\cdot\mbox{\boldmath $\nabla$}){\rm\bf v}=
	-\frac{1}{m}\mbox{\boldmath $\nabla$}(V+Q), 
\end{eqnarray}
where $V$ is the potential appearing in the Schr\"odinger equation and Q is the so-called {quantum potential}. In the presence of a vector potential ${\rm\bf A}$ the (hydrodynamic) velocity ${\rm\bf v}$ is given by
\begin{eqnarray} \lab{v}
	{\rm\bf v}=\frac{1}{m}(\mbox{\boldmath$\nabla$}S-q {\rm\bf A}).
\end{eqnarray}

We see that the chemical potential $\mu_{1}$ in Eq.~(\ref{mu1}) is equivalent to the quantum potential $Q$  in Eq.~(\ref{vQ}).

We consider now the derivation of  Eq.~(\ref{Lorentzforce}). We apply the operator $\mbox{\boldmath$\nabla$}$ to both members of Eq.~(\ref{real}) and use the identities:
\begin{eqnarray}
	\mbox{\boldmath$\nabla$}\frac{m v^2}{2}
	= \, m ({\rm\bf v}\cdot\mbox{\boldmath$\nabla$}) {\rm\bf v}
	+ m {\rm\bf v} \times\left(\mbox{\boldmath$\nabla$}\times{\rm\bf v} \right), 
\end{eqnarray}
\begin{eqnarray}
	\frac{d{\rm\bf v}}{dt}
	\equiv\frac{\partial{\rm\bf v} }{\partial t}+({\rm\bf v} \cdot\mbox{\boldmath$\nabla$}) {\rm\bf v}
	=\frac{\partial{\rm\bf v} }{\partial t}
	+\mbox{\boldmath$\nabla$}\frac{v^2 }{2}
	-{\rm\bf v} \times\left(\mbox{\boldmath$\nabla$}\times{\rm\bf v} \right). 
\end{eqnarray}
where $v^2 = {\rm\bf v} \cdot {\rm\bf v}$. The rotor of Eq.~(\ref{v}) gives
\begin{eqnarray}\lab{rotv}
	\mbox{\boldmath$\nabla$}\times{\rm\bf v}
	=-\frac{q}{m}\mbox{\boldmath$\nabla$}\times{\rm\bf A}
	=-\frac{q}{m}{\rm\bf B}.
\end{eqnarray}
Combining these relations we get Eq.~(\ref{Lorentzforce}).

\section{}\lab{B}

By adopting the notation of Ref. \cite{Barybin}, we put
\begin{eqnarray}\lab{B1}
	\tau_{\rm GL}(T)~\equiv 
\frac{\hbar\gamma(T)}{|\alpha(T)|}, \qquad \quad
	\xi^2_{\rm GL}(T)~\equiv~ \frac{\hbar^2}{2 m |\alpha(T)|},  \qquad \quad
	\eta_{\rm GL}(T)~\equiv~ \frac{\beta(T)}{|\alpha(T)|}~
	\equiv ~\frac{1}{N^{\circ} (T)} ,
\end{eqnarray}
and using Eq.~(\ref{SGL}) and
\begin{eqnarray}
	\hat{R} ~ \equiv ~ \xi^2_{\rm GL}\left(\mbox{\boldmath$\nabla$}+\frac{q}{i\hbar}{\rm\bf A}\right)^2\!
	-\eta_{\rm GL}|\sigma|^2+1,
\end{eqnarray}
Eq.~(\ref{SGL12}) takes the form:
\begin{eqnarray} \lab{R}
	i\hbar\frac{\partial \sigma}{\partial t} = \hat{H}\sigma
	+ {i\hbar}\frac{\hat{R}}{\tau_{\rm GL}}\,\sigma.
	\label{TDGL.eq2}
\end{eqnarray}

Use of the relations
\begin{eqnarray}
\frac{{\rm\bf v} \cdot \mbox{\boldmath$\nabla$} \rho^{1/2}}{\rho^{1/2}} =
\frac{\mbox{\boldmath$\nabla$} \cdot N^{1/2} {\rm\bf v}}{N^{1/2}}~, \qquad \mbox{\boldmath$\nabla$}
\cdot {\rm\bf v}=0 ~.
\end{eqnarray}
then leads us to Eqs.~(\ref{continuity}) and  (\ref{chempot}).

The (stationary) case $\tau_{\rm GL} \rightarrow 0$ (i.e. $\ga \rightarrow 0$), can be studied by considering Eq.~(\ref{SGL12}). Multiplying both members times $\ga$, we see that the limit $\ga \rightarrow 0$ leads to the stationary GL equation (\ref{SGL1}). On the other hand,
by multiplying both members of Eq.~(\ref{R}) times $\tau_{\rm GL}$, the limit $\tau_{\rm GL} \rightarrow 0$ leads to
\begin{eqnarray}
	\hat{R}\sigma \equiv -\left(G-i\xi^2_{\rm GL}\frac{m}{\hbar}
	\frac{\mbox{\boldmath$\nabla$}\cdot{\rm\bf J} }{\rho }\right)^2\sigma
	= 0
	\label{stationaryGL}
\end{eqnarray}
The real part of Eq.~(\ref{stationaryGL}) gives $G \, \si = 0$, i.e.
\begin{eqnarray}\lab{chi}
	\xi^2_{\rm GL}\mbox{\boldmath$\nabla$}^2\chi
	+\left[1-\xi^2_{\rm GL}\left(\frac{m v }{\hbar}\right)^2\right]\chi
	-\chi^3 =0,
\end{eqnarray}
where $\chi=|\sigma|/|\sigma^{\circ}|\equiv(N /N^{\circ} )^{1/2}$.
The vanishing of the imaginary part of Eq.~(\ref{stationaryGL}) implies
\begin{eqnarray}
	\mbox{\boldmath$\nabla$}\cdot{\rm\bf J}=0, \qquad {\rm i.e. } \qquad {\rm\bf v} \cdot\mbox{\boldmath$\nabla$}N =0,
\end{eqnarray}
which means that in the stationary case non-homogeneities in $N$ only arise in the  plane perpendicular to ${\rm\bf v}$.

\section{}\lab{C}

We can define $\mu ~ \equiv ~ \mu_{0}+\mu_{1}+\mu_{2}$ and introduce the total chemical potential (cf. Eqs.~(\ref{chempot}) and (\ref{chempot2}))
\begin{eqnarray}
	\mu_{\Sigma} \,\,
	= \mu +\frac{mv^2 }{2}
	~ \equiv ~\!\!\!\left(\mu_{0}-\frac{\hbar^2}{2m}
	\frac{\mbox{\boldmath$\nabla$}^2N^{1/2} }{N^{1/2} }
	-\frac{\xi^2_{\rm GL}}{\tau_{\rm GL}}\frac{m}{q}
	\frac{\mbox{\boldmath$\nabla$}\cdot{\rm\bf J} }{N }\right)
	+\frac{mv^2 }{2}.
\end{eqnarray}
The gauge invariant potential $\Phi$ is defined as
\begin{eqnarray}
	\Phi\!\!\!~ &\equiv& ~ \!\!\! q \varphi+\hbar \frac{\partial \theta}{\partial t}
	=-\left(\mu +\frac{mv^2 }{2}\right)
	\equiv -\mu_{\Sigma}.
\end{eqnarray}
and we see that $\mu_{\Sigma}$ is always opposite in phase to $\Phi$. We may also introduce  $\zeta  \equiv \mu + q \varphi$ and the electrochemical potential $\zeta_{\Sigma}$   given by
\begin{eqnarray}
	\zeta_{\Sigma}\!\!\! ~ &\equiv& ~\!\!\! \zeta +\frac{mv^2 }{2}
	\equiv ~ \mu_{\Sigma}+q \varphi
	=-\hbar \frac{\partial \theta}{\partial t}.
\end{eqnarray}
We see that it depends on the rate of change of the NG mode (the {\it phason} mode $\theta$). Finally,  the gauge invariant vector and scalar potentials are given by
\begin{eqnarray}
	{\rm\bf A}_{\rm g}=-\frac{m{\rm\bf v} }{q}
	\qquad {\rm and} \qquad
	\varphi_{\rm g}=\frac{\Phi}{q},
\end{eqnarray}
respectively. The corresponding electric field is
\begin{eqnarray}
	{\rm\bf E}=-\frac{\partial {\rm\bf A}_{\rm g}}{\partial t}
	-\mbox{\boldmath$\nabla$}\varphi_{\rm g}
\end{eqnarray}
The Coulomb gauge condition $\mbox{\boldmath$\nabla$} \cdot {\rm\bf A}_{\rm g} = 0$  then implies that
\begin{eqnarray}\lab{divv}
	\mbox{\boldmath$\nabla$}\cdot{\rm\bf v} =0
\end{eqnarray}
and Eq.~(\ref{rotv}) is also reobtained.
Due to Eq.~(\ref{divv}), the velocity normal to any boundary thus vanishes, ${\rm\bf n}\cdot{\rm\bf v} =0$, and at boundaries any current such that ${\rm\bf n}\cdot{\rm\bf v} \neq 0$ converts into a current with  ${\rm\bf n}\cdot{\rm\bf v} =0$ and the condensate density $N$ has nonuniform distribution.


\section*{References}


\begin{thebibliography}{00}



\bibitem{Barham} J. Barham,
Biosystems {\bf 38},  235 
(1996)



\bibitem{Freeman2008} W. J.~Freeman,
Neural Networks {\bf 21}, 257 
(2008)\\ http://repositories.cdlib.org/postprints/2781

\bibitem{11a} W. J. Freeman and G. Vitiello,
J. Phys. A: Math. Theor. {\bf 41},  304042 (2008);
http://Select.iop.org; q-bio.NC/0701053v1

\bibitem{Freeman2010} W. J.~Freeman and G.~Vitiello,
Int. J. Mod. Phys. B {\bf 24}, 3269 
(2010) \\ http://dx.doi.org/10.1142/S0217979210056025

\bibitem{Freeman2001} W. J. ~Freeman, {\it How Brains Make Up Their
Minds},   Columbia UP, New York 2001


\bibitem{Braitenberg} V.~ Braitenberg and A.~Sch\"uz,
{\it Cortex: Statistics and Geometry of Neuronal Connectivity}, 2nd edn. Springer-Verlag, Berlin  1998

\bibitem{11} W. J. Freeman and G. Vitiello,
Phys. of Life Reviews {\bf 3}, 93 (2006); \\
http://dx.doi.org/10.1016/j.plrev.2006.02.001, http://repositories.cdlib.org/postprints/1515,
q-bio.OT/0511037


\bibitem{Freeman19} W.J. Freeman and G. Vitiello,
J. Physics Conf Series {\bf 174}, 012011 (2009).
http://www.iop.org/EJ/toc/1742-6596/174/1.

\bibitem{9} W. J. ~Freeman,
Clin. Neurophysiol. {\bf 115}, 2077 
 (2004);\\ http://repositories.cdlib.org/postprints/1006

\bibitem{10} W. J. Freeman,
Clin. Neurophysiol. {\bf 115}, 2089 
(2004);\\ http://repositories.cdlib.org/postprints/1486

\bibitem{12} W. J. Freeman,
Clin. Neurophysiol. {\bf 116} (5), 1118 
(2005); \\ http://repositories.cdlib.org/postprints/2134,
http://authors.elsevier.com/sd/article/S1388245705000064

\bibitem{13} W. J. Freeman,
Clin. Neurophysiol.  {\bf 117} (3), 572 
(2006);\\ http://repositories.cdlib.org/postprints/1480/,
  http://dx.doi.org/10.1016/j.clinph.2005.10.025


\bibitem{Bollobas} W. J.~ Freeman, R.~ Kozma, B.~ Bollob\'as, O. Riordan,
Chapter 7. {\it Scale-free cortical planar network}, in: {\it Handbook of Large-Scale Random Networks. Series: Bolyai Mathematical Studies}, Vol. 18, B.~ Bollob\'as, R.~ Kozma, D.~ Mikl\"os  (Eds.), Springer, New York 2009,  pp. 277 

\bibitem{Vitiello:1995wv}
G.~Vitiello,
Int.\ J.\ Mod.\ Phys.\ B {\bf 9},  973 (1995).

\bibitem{Vitiello:2001} G.~Vitiello,
{\it My Double Unveiled}, John Benjamins, Amsterdam,   2001.

\bibitem{Alfinito:2002a} E. Alfinito, O. Romei and G. Vitiello, Mod.
Phys. Lett. B {\bf  16}, 93 (2002).

\bibitem{Alfinito:2002b} E. Alfinito and G. Vitiello, Phys. Rev.
 B {\bf  65}, 054105 (2002).

\bibitem{Kitzbichler} M. G.~Kitzbichler, M. L.~Smith, S. R.~Christensen, E.~
Bullmore,
PLoS Comput Biol {\bf 5} (3), e1000314 (2009).\\ doi:10.1371/journal.pcbi.1000314


\bibitem{DifettiBook} M.~Blasone, P.~Jizba and G.~Vitiello, {\sl Quantum Field Theory and its macroscopic manifestations}, Imperial College Press, London,  2011

\bibitem{Umezawa:1993yq} H.~Umezawa, {\sl Advanced field theory: Micro, macro,
and thermal physics}, AIP, New York 1993.

\bibitem{DelGiudice:1985}
E. Del Giudice, S. Doglia, M. Milani and G. Vitiello,
Nucl. Phys. B {\bf 251} (FS 13), 375  (1985).

\bibitem{DelGiudice:1986}
E. Del Giudice, S. Doglia, M. Milani and G. Vitiello, Nucl. Phys. B
{\bf 275} (FS 17), 185 (1986).


\bibitem{Goldstone:1961eq} J. Goldstone,
     Nuovo Cimento {\bf 19},   154  (1961).\\
J. Goldstone, A. Salam and S. Weinberg, Phys. Rev. {\bf 127}, 965
(1962).

\bibitem{ITZ}  C.~Itzykson and J.~Zuber, {\it Quantum
field theory}, McGraw-Hill, New York,   1980.


\bibitem{Umezawa:1982nv} H.~Umezawa, H.~Matsumoto and M.~Tachiki,
     {\sl Thermo Field Dynamics and condensed states},
     North-Holland, Amsterdam, The Netherland 1982.

\bibitem{Anderson:1984a}
P.W. Anderson,
{\it Basic Notions of Condensed Matter Physics}, Benjamin, Menlo
Park,  1984.


\bibitem{hig} P. Higgs,
{Phys. Rev.}   {\bf 145}, 1156 (1966).\\
T.W.B. Kibble,
Phys. Rev.  {\bf 155},  1554 (1967).

\bibitem{Leplae} L.~Leplae and H.~Umezawa,
Nuovo Cim. {\bf 44}, 410 
(1996)

\bibitem{Matsumoto:1975fi} H. Matsumoto, N. J. Papastamatiou, H. Umezawa
and G. Vitiello,
Nucl. Phys. B {\bf 97}, 61 (1975).

\bibitem{Matsumoto:1975rp} H. Matsumoto, N. J. Papastamatiou, H. Umezawa, Nucl. Phys. B {\bf 97}, 90
(1975).

\bibitem{Barybin} A.A. Barybin,
Advances in Condensed Matter Physics, 425328 (2011)

\bibitem{Haken:1984a} H. Haken,
in {\it Proc.Int. School of Physics E.Fermi, Nonlinear
spectroscopy},
ed. N. Bloembergen, North-Holland, Amsterdam 1977, p.350\\
\noindent H. Haken, {\it Laser theory}, Springer-Verlag, Berlin
1984



\bibitem{Bettencourt} L.M.A.~Bettencourt, N.A.~Antunes and W.H.~Zurek,
Phys. Rev. {\bf D62}, 065005 (2000).\\
N.A.~Antunes, L.M.A.~Bettencourt  and W.H.~Zurek,
Phys. Rev. Lett. {\bf 82}, 2824 (2000).

\bibitem{Crooks} As a general references see
G. E.~Crooks,
J. Stat. Phys. 90, 1481 (1998).\\
C. ~Jarzynski,
Phys. Rev.
Lett. 78, 2690 (1997).\\
C. ~Jarzynski,
Phys. Rev. E 56, 5018 (1997).



\bibitem{Manka:1990} R.~Manka and G.~Vitiello,
Annals of Phys. {\bf 199}, 61 (1990)


\bibitem{Vitiello:2004}
G.~Vitiello, Int. J. Modern Physics B {\bf 18},   785 (2004).

\bibitem{Pessa2003} E.~Pessa   and G.~Vitiello,
Mind and Matter {\bf 1},  59 (2003).


\bibitem{Pessa:2004}
E. Pessa   and G. Vitiello,
Int. J. Modern Physics B {\bf 18},   841 (2004).


\bibitem{Abbot} B. C.~Abbot,
Journal of General Physiology {\bf 43},
119 
(1960)



\bibitem{kib2} T.W.B. Kibble,
in  {\it Topological defects and the non-equilibrium dynamics of
symmetry breaking phase transitions}, eds. Y.M. Bunkov and H.
Godfrin, NATO Science Series C 549, Kluwer Acad., Dordrecht,  2000,
p. 7.\\
G.E. Volovik,
in {\em Topological defects and the non-equilibrium dynamics of
symmetry breaking phase transitions}, eds. Y.M. Bunkov and H.
Godfrin, NATO Science Series C 549, Kluwer Acad., Dordrecht,  2000,
p. 353.


\bibitem{zurek1} W.H. Zurek,
Phys. Rep.  {\bf 276}, 177 (1997) and Refs. therein quoted.


\bibitem{38} R. Kozma, Unpublished data (2005).

\bibitem{Perelomov:1986tf}
A.M.~ Perelomov, {\it Generalized Coherent States and their
Applications}, Springer, Berlin,   1986.

\bibitem{FractalsNMNC} G. Vitiello,
New Mathematics and Natural Computing {\bf 5}, 245 (2009).

\bibitem{FractalQI} G. Vitiello, Fractals and the Fock-Bargmann
representation of coherent states, in {\it Quantum Interaction. Third International Symposium (QI-2009)},
Saarbruecken, Germany, Eds. P. Bruza, D. Sofge, et al.. Lecture Notes in Artificial Intelligence, Edited by R.Goebel, J. Siekmann, W.Wahlster, Springer-Verlag Berlin Heidelberg 2009, (pp. 6-16).

\bibitem{Peitgen} H.O. Peitgen, H. J\"urgens and D. Saupe, {\it Chaos and
fractals. New frontiers of Science}, Springer-Verlag, Berlin 1986

\bibitem{BakinBunde} P. Bak and M Creutz,
in {\it Fractals in Science},  eds. A. Bunde and S. Havlin,
Springer -Verlag, Berlin 1995.




\bibitem{Freeman1975} W. J.~Freeman, {\it Mass Action in the Nervous System}, Academic Press,  New York 1975, 	© 2004.\\ http://sulcus.berkeley.edu/MANSWWW/MANSWWW.html

\bibitem{Freeman14}
W.J. Freeman and J. Zhai,
Cogn. Neurodyn. {\bf 3} (1), 97 
(2009).\\ http://repositories.cdlib.org/postprints/3374, http://dx.doi.org/10.1007/s11571-008-9064-y

\bibitem{Miller} L.M.~ Miller and C.E.~Schreiner
J Neurosci {\bf 20}, 7011 
(2000)


\bibitem{Skarda} C.A.~Skarda, W.J.~Freeman
Brain and Behavioral Science {\bf 10}, 161
(1987)

\bibitem{FreemanErwin}  W. J.~Freeman and H.~Erwin,
{\it Freeman K-set}. Scholarpedia {\bf 3}(2),  3238 (2008).\\
http://www.scholarpedia.org/article/Freeman K-set

\bibitem{Rice1950} S.O. Rice, {\it Mathematical Analysis of Random Noise and
Appendixes}. Technical Publications Monograph B-1589. Bell Telephone
Labs Inc., New York 1950.

\bibitem{FreemanCognitive} W.J. Freeman, Cognitive Neurodynamics {\bf 3}
(1),  105
(2009).\\
http://repositories.cdlib.org/postprints/3387.

\bibitem{Principe} J.C.~Principe,  V.G.~ Tavares, J.G.~Harris and W. J.~Freeman,
Design and implementation of a biologically realistic olfactory cortex in analog VLSI.  Proceedings IEEE {\bf 89},  1030 
(2001)


\end{thebibliography}
\end{document}